\documentclass[useAMS,usenatbib]{mn2e}

\usepackage{graphicx,color}
\usepackage[usenames,dvipsnames]{xcolor}
\usepackage{times}
\usepackage{natbib}
\usepackage{setspace}
\newif\ifAMStwofonts
\AMStwofontstrue

\newcommand{\hMsun}{$\,h^{-1}{\rm M}_\odot$}
\newcommand{\hMpc}{\mbox{$h^{-1}{\rmn{Mpc}}~$}}
\newcommand{\hkpc}{\mbox{$h^{-1}{\rmn{kpc}}~$}}

\newcommand{\msun}{M$_\odot$}
\newcommand{\mor}{{\it morgana}}
\newcommand{\del}{{\it delucia}}
\newcommand{\gal}{{\it galacticus}}
\newcommand{\mtwoh}{{$M_{200}$}}
\newcommand{\mfof}{{$M_{\rm FoF}$}}

\newcommand{\mincir}{\raise
  -2.truept\hbox{\rlap{\hbox{$\sim$}}\raise5.truept \hbox{$<$}\ }}
\newcommand{\magcir}{\raise
  -2.truept\hbox{\rlap{\hbox{$\sim$}}\raise5.truept \hbox{$>$}\ }}
\newcommand{\siml}{\raise
  -2.truept\hbox{\rlap{\hbox{$\sim$}}\raise5.truept \hbox{$<$}\ }}
\newcommand{\simg}{\raise
  -2.truept\hbox{\rlap{\hbox{$\sim$}}\raise5.truept \hbox{$>$}\ }}

\title[Testing cooling models against hydrodynamical simulations]
{A semi-analytic model comparison: testing cooling models against
  hydrodynamical simulations}

\author[P. Monaco et al.] {
Pierluigi Monaco$^{1,2}$\thanks{Email: monaco@oats.inaf.it}, Andrew
J. Benson$^3$, Gabriella De Lucia$^2$, Fabio Fontanot$^{4,5}$, \and 
Stefano Borgani$^{1,2,6}$, Michael Boylan-Kolchin$^{7,8}$ \\ 
$^1$ Dipartimento di Fisica - Sezione di Astronomia, Universit\`a di Trieste, via Tiepolo 11, I- 34131 Trieste, Italy\\
$^2$ INAF, Osservatorio Astronomico di Trieste, Via Tiepolo 11, I-34131 Trieste, Italy \\
$^3$ The Observatories of the Carnegie Institution for Science, 813 Santa Barbara St, Pasadena, CA 91101, USA\\
$^4$ Heidelberger Institut f\"ur Theoretische Studien (HITS), Schloss-Wolfsbrunnenweg 35, 69118 Heidelberg, Germany\\
$^5$ Institut f\"ur Theoretische Physik, Philosophenweg 16, 69120, Heidelberg, Germany\\
$^6$ INFN-National Institute for Nuclear Physics, Via Valerio 2, I-34127 Trieste, Italy\\
$^7$ Center for Cosmology, Department of Physics and Astronomy, 4129 Reines Hall, University of California, Irvine, CA 92697, USA\\
$^8$ Department of Astronomy, University of Maryland, College Park, MD 20742, USA
}

\begin{document}

\maketitle

\label{firstpage}

\begin{abstract}

We compare predictions of cooled masses and cooling rates from three
stripped-down Semi-Analytic Models (SAMs) of galaxy formation with the
results of N-body+SPH simulations with gas particle mass of $3.9
\times 10^6$ {\hMsun}, where radiative cooling of a gas of primordial
composition is implemented.  We also run a simulation where cooling is
switched on at redshift $\sim$2, in order to test cooling models in a
regime in which their approximations are expected to be valid.  We
confirm that cooling models implemented in SAMs are able to predict
the amount of cooled mass at $z=0$ to within $\sim$20 per cent.
However, some relevant discrepancies are found. (i) When the
contribution from poorly resolved halos is subtracted out, SAMs tend
to under-predict by $\sim$30 per cent the mass that cools in the
infall-dominated regime. (ii) At large halo masses SAMs tend to
over-predict cooling rates, though the numerical result may be affected
by the use of SPH.  (iii) As found in our previous work, cooling rates
are found to be significantly affected by model details: simulations
disfavour models with large cores and with quenching of cooling at
major mergers.  (iv) When cooling is switched on at $z\sim2$, cold gas
accumulates very quickly in the simulated halos. This accumulation is
reproduced by SAMs with varying degrees of accuracy.

\end{abstract}


\section{Introduction}
\label{introduction}

The formation of galaxies within the $\Lambda$CDM cosmological model
involves a large number of physical processes, many of which are still
poorly understood.  The hierarchical build-up of DM halos, resulting
from the non-linear evolution of primordial perturbations under their
own gravity, provides the backbone of the whole process of galaxy
formation.  Thanks to advances in N-body techniques and to the very
accurate constraints available on cosmological parameters
\citep[e.g.,][]{Planck13}, the evolution and properties of DM halos
can be computed with very good accuracy \citep[e.g.][]{Reed13}.
Despite baryonic processes are known to affect the build-up of DM
halos to some extent \citep[e.g.][]{Stanek09, Duffy10, Saro10, Cui12},
the modeling of baryonic physics still provides the major source of
uncertainty.  A purely collisionless simulation thus remains a good
starting point for a galaxy formation model.

Galaxy formation in the cosmological context has been historically
addressed with two main tools.  SAMs are applied to the backbone of DM
halo merger trees, taken from an N-body simulation or equivalent tools
\citep[e.g.][]{Monaco13}.  They use a set of simplified or
phenomenological models to describe the various processes that involve
baryons.  Hydrodynamic simulations consist of numerically solving the
equations of motion of DM and gas particles in a realization of a
cosmological volume.  Two fundamental processes take place on scales
that are within reach of presently available simulations.  Gravity and
hydrodynamic forces are responsible for infalling of gas and heating
to the halo virial temperature.  Radiative cooling down to $\sim10^4$
K, which is computed based on the density and temperature of the
heated gas, is responsible for the condensation of gas into the inner
regions of the halo. There, the gas can fragment into stars and thus
form a galaxy, but star formation and all the processes triggered by
it (stellar feedback, chemical evolution, galaxy winds, black hole
seeds), not to mention the accretion of gas onto black holes, take
place at much smaller scales. To properly follow these processes, the
range of scales that must be resolved (from sub-pc scale of star
formation to cosmological scales) is so vast and the involved physics
so complex that their effects need to be treated through simplified,
sub-resolution models.

The modeling of cooling in SAMs is based on the assumption that the
gas settles into a hot atmosphere in hydrostatic equilibrium within
the potential well of the DM halo.  This allows the computation of a
cooling time as a function of radius.  Whenever the central cooling
time is longer than the halo dynamical time \citep[``cooling-dominated
  regime'';][]{Rees77,Binney77,White91}, the deposition of cold gas
into the central galaxy is assumed to be regulated by cooling;
otherwise, the timescale for gas to condense into the galaxy is
assumed to be of the order of the halo dynamical time
(``infall-dominated regime'').  Using hydrodynamical simulations that
included radiative cooling and star formation (but no efficient
stellar feedback), \cite{Keres05} reported that at $z\ga2$, or at any
redshift for halos smaller than $\sim10^{11}$ {\msun} \citep[see
  also][]{Dekel06}, gas tends not to shock to the virial temperature
but to condense directly into the galaxy via a cold flow.  As a
caveat, the deposition of gas through cold flows is known to depend on
the hydrodynamic scheme \citep{Nelson13} and to be affected by
feedback processes connected to galaxy formation
\citep{Benson11,Murante12}.

To check to what level the two techniques, SAMs and hydrodynamical
simulations, give a consistent description of the deposition of cold
mass into the ``galaxies'', many authors
\citep{Benson01,Yoshida02,Helly03,Cattaneo07,Viola08,Saro10,Lu11,Hirschmann12}
performed comparisons of the predictions of SAMs and simulations.  All
of these papers presented comparisons performed using stripped-down
SAMs, where all processes beyond gas cooling were switched off and
simulations where either no star formation or no (effective) feedback
from star formation was present.  
The first papers
\citep{Benson01,Yoshida02,Helly03,Cattaneo07} compared one SAM with
simulations and reported that SAMs are able to reproduce the gas mass
that cools in DM halos to a level, to cite \cite{Benson01}, {\it
  ``better than a pessimist might have expected''}.  More recent
papers focused on some discrepancies between the two kinds of
modeling.  \cite{Viola08} simulated cooling in isolated, hydrostatic
DM halos and compared the resulting cooling mass with two models: an
implementation from \cite{Cole00} and the one used in the MOdel of the
Rise of Galaxies And Agn \citep[MORGANA][]{Monaco07}.  They found that
the former model underestimates the amount of cooled mass when cooling
is suddenly switched on, while the latter model produced a much better
fit.  \cite{Saro10} compared the galaxy populations in a massive
galaxy cluster predicted by SPH simulations with those from the SAM
described in \cite{DeLucia07}. Both models included gas cooling and a
simple prescription for star formation (as we will do in this study).
The resultant object-by-object comparison revealed important
differences between the two methods. In particular, the star formation
history of BCGs in the SPH simulations is characterized by a more
prominent high redshift peak and lower level of recent star formation
with respect to predictions from the SAM. As noticed by \cite{Saro10},
this is due to the assumption of an isothermal gas density
distribution for the hot gas in the SAM, which differs from the actual
gas distribution in the simulation.

In \citet[][hereafter Paper I]{DeLucia10} we compared the results of
stripped-down versions of three independently developed SAMs,
``Durham'' \citep{Cole00,Benson01}, ``Munich''
\citep{DeLucia07,Saro10} and Morgana \citep{Monaco07}.  We ran them on
the same set of merger trees taken from the Millennium simulation
\citep{Springel05}.  We concentrated on two mass scales, selecting 100
halos as massive as the Milky Way at $z=0$ and 100 as abundant as
SCUBA galaxies at $z\sim2$.  We found that the resulting cold masses
and cooling rates were in good agreement at the Milky Way mass scale,
but we noticed that the ``Durham'' model predicted systematically
lower cooling rates in halos of the SCUBA set.  We showed that this
difference is again due to different assumptions on the gas profile,
which is isothermal in the ``Munich'' model and cored in the
``Durham'' model.  A comparison of the results of the same models in a
configuration where star formation and stellar feedback is active was
presented in \cite{Fontanot13}.

More recently, \cite{Lu11} compared the results of several
implementations of semi-analytic cooling models with 1D simulations of
both an isolated halo and an accreting halo, and with a simulation of a
cosmological volume including inefficient stellar feedback.  They
reported that different models give predictions of gas accretion rates
that can vary by up to a factor of 5. With respect to simulations,
semi-analytic models under-predict gas accretion rates in small halos
and over-predict them in massive halos.  They also showed that the
predicted cooling-dominated and infall-dominated regimes do not
closely correspond to the regimes where cold flows or hot atmospheres
were found to dominate in their simulation.  
A similar setting was used in \cite{Hirschmann12}, with the aim of
comparing the predictions of simulations and SAMs when feedback and
star formation are used.  They used a set of resimulations of
dark-matter halos \citep[see][]{Oser10} run with the {\sc gadget}
code, but they implemented primordial cooling and inefficient stellar
feedback, and compared the results with several versions of the model
of \cite{Somerville08}, including a stripped-down one with no stellar
feedback.  This work shows that, while stripped-down SAMs and
simulations agree rather well (with some discrepancies consistent with
those found in the papers cited above), SAMSs predict that the
deposition of cold gas takes place mostly in the cooling-dominated
regime, while the contribution from simulated cold flows is
significant at all redshifts and dominant at $z\ga1-2$.

To perform an accurate comparison of simulations with SAMs, the SAMs
should be run on merger trees extracted from the same simulation, as
was done in \cite{Yoshida02}, \cite{Helly03}, \cite{Cattaneo07},
\cite{Saro10} and \cite{Hirschmann12} but not in \cite{Benson01} and
\cite{Lu11}.  Proper time sampling of merger trees is also relevant:
\cite{Benson12} demonstrated that a SAM gives a convergent description
of the formation of galaxies if merger trees are sampled at least
$\sim128$ times, a factor of two higher than, e.g., the time sampling
of the Millennium simulation.  Finally, the condensation of gas into
the central ``galaxy'' is governed not only by cooling but also by the
time required for progenitor halos to merge with the central object.
As demonstrated in Paper I, different SAMs do not use consistent
predictions of galaxy merging times, and this should be properly taken
into account.

A further important issue is related to the runaway nature of cooling
of self-gravitating gas.  When no feedback from stars is present, gas
(over-)cools in the small DM halos that form at high redshift, so
cooling takes place in poorly resolved halos, and in the
infall-dominated regime where the cooling models we are aiming to test
do not apply.  Increasing the resolution would imply 
resolving smaller halos at higher redshift,
thus worsening the over-cooling problem.  Results showing a reasonable
agreement of SAMs and simulations at $z=0$ may simply reflect the fact
that both predict that most baryons have (over-)cooled.

We are convinced that
a satisfactory numerical test of semi-analytic cooling
models is still missing.  An ideal test should have the
following characteristics: (i) the study should be limited to cooling
in well resolved halos; (ii) a clear assessment of the accumulation of
cold gas in the cooling-dominated regime, independent 
of (over-)cooling at high redshift, should be performed; (iii)
the analysis should include several SAMs; (iv) SAMs should
be run on merger trees extracted from the same simulation; (v) time
sampling should be fine enough to guarantee convergence of SAM
predictions; (vi) measured cooling rates should be independent of the
accuracy with which merger times are predicted by SAMs.
In this paper we present a comparison of SAMs and numerical (SPH)
simulations that meets all the above criteria.

Finding a numerical solution of cooling in cosmological halos is not
a straightforward task.  To face this problem we use the 
{\sc gadget} Tree-PM$+$SPH code.  
The specific implementation used in this paper is shortly
described in Section~\ref{section:code}.  While this project was in
progress, \cite{Keres12} published a comparison of {\sc gadget} with
the new moving-mesh code {\sc arepo} \citep{Springel10}, run on the
same initial conditions and with a setting similar to that of \cite{Lu11},
though at higher resolution.
They reported that galaxy stellar mass functions obtained with the two
hydrodynamic schemes tend to be quite similar in the low-mass end but
exhibit significant differences in the high mass end.  They traced the
origin of this difference to the different efficiencies of dissipative
heating from gas accretion onto halos, which cause cooling to be
partially offset in SPH with respect to the moving--mesh hydrodynamic
scheme.  The issue was readdressed by
\cite{Nelson13}, who showed that the strucuture of cold filaments
penetrating into hot halos is very different for the two hydro solvers.
While a detailed comparison of different hydrodynamical schemes is
beyond the aims of this paper, we stress that some of our results
might be affected by the use of SPH. We will comment on this below.

As in Paper I, we use here three stripped-down SAMs.  The model of
\cite{DeLucia07}, denoted the {\it Munich} model in Paper I, will be
called here {\del}, while {\mor} of \cite{Monaco07} will retain its
acronym.  In place of the {\it Durham} model we use the highly modular
{\gal} model of \cite{galacticus} in a configuration that closely
resembles that of \cite{Bower06}.  We run the three SAMs on merger
trees extracted from a collisionless N-body simulation, and compare
them with the results of a simulation run on the same initial
conditions but including gas hydrodynamics and radiative cooling for a
primordial composition.
We set merger times to zero in SAMs, and compute cooled masses 
in simulations by summing 
over all substructures in simulated FoF halos.
We run the hydrodynamical simulation in two
configurations: we allow cooling to be active from the start, or we
run the simulation without radiative cooling down to redshift
$z\sim2$, and then switch cooling on.  This second configuration
allows us to test cooling models exactly in the range of redshift and
halo mass where their approximations are expected to be valid: 
when cooling is
switched on,
no (over-)cooling has taken place in the infall-dominated regime and
all the
baryons associated with halos are in hot atmospheres;
the contribution of cooling from small halos (that
are poorly resolved and where gas mostly cools in the infall-dominated
regime) is much smaller.

We use a relatively small box (36 \hMpc on a side, 50 Mpc for
$h=0.72$) sampled with $512^3$ DM particles and an equal number of gas
particles, and use a force resolution of 1.5 comoving \hkpc.  With
these choices, we prioritize resolution over statistics and test for
the first time cooling at a resolution that is sufficient to resolve
the morphology of a galaxy when star formation and feedback are
properly taken into account \citep[e.g.][]{Scannapieco12}.  The DM
halo masses we test range from small galaxies ($3\times10^{11}\ {\rm
  M}_\odot$) to rich galaxy groups ($5\times10^{13}\ {\rm M}_\odot$).
Merger trees have been sampled at 128 time-steps \citep[uniformly in
  the log of scale factor;][]{Benson12}, with the time interval
between two snapshots roughly corresponding to half the dynamical time
of DM halos.  Moreover, following \cite{Viola08} and \cite{Saro10} we
use a ``star formation'' algorithm to remove cooled gas particles from
hydrodynamics, with the result of speeding up the simulation and
strongly reducing numerical artifacts at the interface of cold and hot
phases.

We will be mainly concerned with cooling rates and cooled masses,
neglecting for the time being whether gas has cooled on the main
substructure of the halo or on a satellite.  Other important issues,
like the density profiles of gas in cooling halos, the quantification
of cooling on non-central galaxies, or the delineation between
cold-flow and hot-flow modes, will be addressed in future works.  This
paper is organized as follows.  Section~\ref{section:sim} describes
the simulations run for this project and the post-processing analysis
used to obtain cooling rates and cooled masses.
Section~\ref{section:sam} presents the three SAMs used in the paper.
Section~\ref{section:results} reports the comparison of results from
simulations and SAMs.  Finally, 
Section~\ref{section:comparison} compares the results
  with those presented in previous papers, and
Section~\ref{section:discussion} gives
a summary of the main results and a discussion.

\begin{figure}
\centering{\includegraphics[width=.45\textwidth]{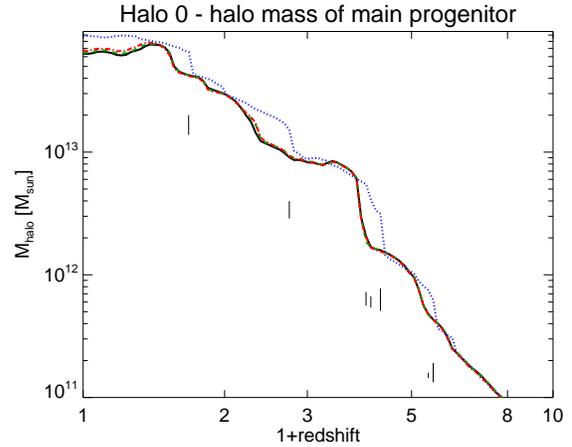}}
\caption{Mass accretion history of the main progenitor of Halo 0, the
  most massive halo in our box.  Black continuous line: {\mtwoh} from
  the PUREDM simulation.  Blue dotted line: {\mfof} from the PUREDM
  simulation.  Green dashed line: {\mtwoh} from the COOL
  simulation. Red dot-dashed line: {\mtwoh} from the COOLZ2
  simulation. Vertical dashes denote major mergers with mass ratios of
  merging halos larger than 1:10, the length being proportional to the
  mass ratio.}
\label{fig:mah0}
\end{figure}

\section{Simulations}
\label{section:sim}

\subsection{The code}
\label{section:code}

The simulations in this paper were performed with 
version 3 of the {\sc gadget} code \citep{Springel05gad}, a massively
parallel TreePM+SPH code with fully adaptive time-step integration.
To describe the hydrodynamical evolution of gas, the code adopts the 
SPH formulation of {\sc gadget2} where entropy is explicitly conserved
\citep{Springel05gad}.  
Whenever a Lagrangian SPH code is used, cooling of hot gas
at the interface with a cold condensation is affected by a numerical
bias, due to the fact that density estimates of hot particles are
affected by the many cold SPH neighbors.  As shown by
\cite{Yoshida02} and \cite{Tornatore03}, the entropy-conserving
formulation of SPH 
limits the importance of this numerical effect.
Another way to limit
it is to avoid the formation of cold clumps using a ``star formation''
recipe to transform cooled particles in collisionless ones
\citep{Viola08,Saro10}; this is described below.
Radiative cooling of a plasma with primordial composition is computed
as in \cite{Katz96}, assuming an ionizing UV background switched on at $z\sim6$
and evolving with redshift as suggested in, e.g., 
\cite{Haardt96}.

In the absence of heating from star formation, following the evolution
of gas subject to runaway cooling is computationally expensive, as
most time is devoted to integrating the hydrodynamics of over-cooled
and unresolved gas condensations with sizes that are much smaller than
the gravitational softening.  Since cooling is irreversible in these
conditions, it is convenient to transform cooled particles into
collisionless ``stars''.  We thus used a ``star formation'' algorithm
where every gas particle with overdensity higher than $\delta_{\rm
  cold}$ (with respect to the average gas density) and temperature
lower than $T_{\rm cold}=10^5$ K is instantaneously transformed into a
collisionless ``star'' particle.  This causes a dramatic speed-up of
the code; moreover, the absence of overcooled gas condensations helps
in reducing numerical overcooling at the interface of cold and hot
gas.  This choice has modest drawbacks: the lack of pressure from the
cold gas will influence the profile of the hot particles in some way
(but in any case the pressure of over-cooled gas blobs is unphysical),
and collisionless particles will slowly scatter into a ``diffuse
light'' component.  

We stress that this ``star formation'' prescription is not meant to
describe the formation of stars but it only represents a convenient
numerical shortcut to treat particles that have undergone runaway
cooling.  In the following, we will call refer to ``star'' particles
as cooled particles.

Using a small 9 {\hMpc} box with $128^3$ DM and $128^3$ gas particles
and with the same mass and force resolution as the simulations
presented here, we performed four test simulations using 
values of $\delta_{\rm cold}=1000$, $10000$, $30000$ and an effectively infinite value to 
reproduce to the case of no star formation.
Comparing the results of the simulations at $z=2.9$ (where the
simulation without star formation slowed down dramatically), we
checked that the density and temperature profiles of hot gas were
reasonably stable for changes in $\delta_{\rm cold}$, with $10000$
being a good value to preserve hot gas profiles while achieving
significant speedup.  Furthermore, we checked that, for the three runs
with star formation, the mass of cooled gas changes by no more than
$\sim10-20$ per cent at all redshifts, higher thresholds giving
slightly lower values of cooled mass as expected.

\begin{figure*}
\centering{\includegraphics[width=.95\textwidth]{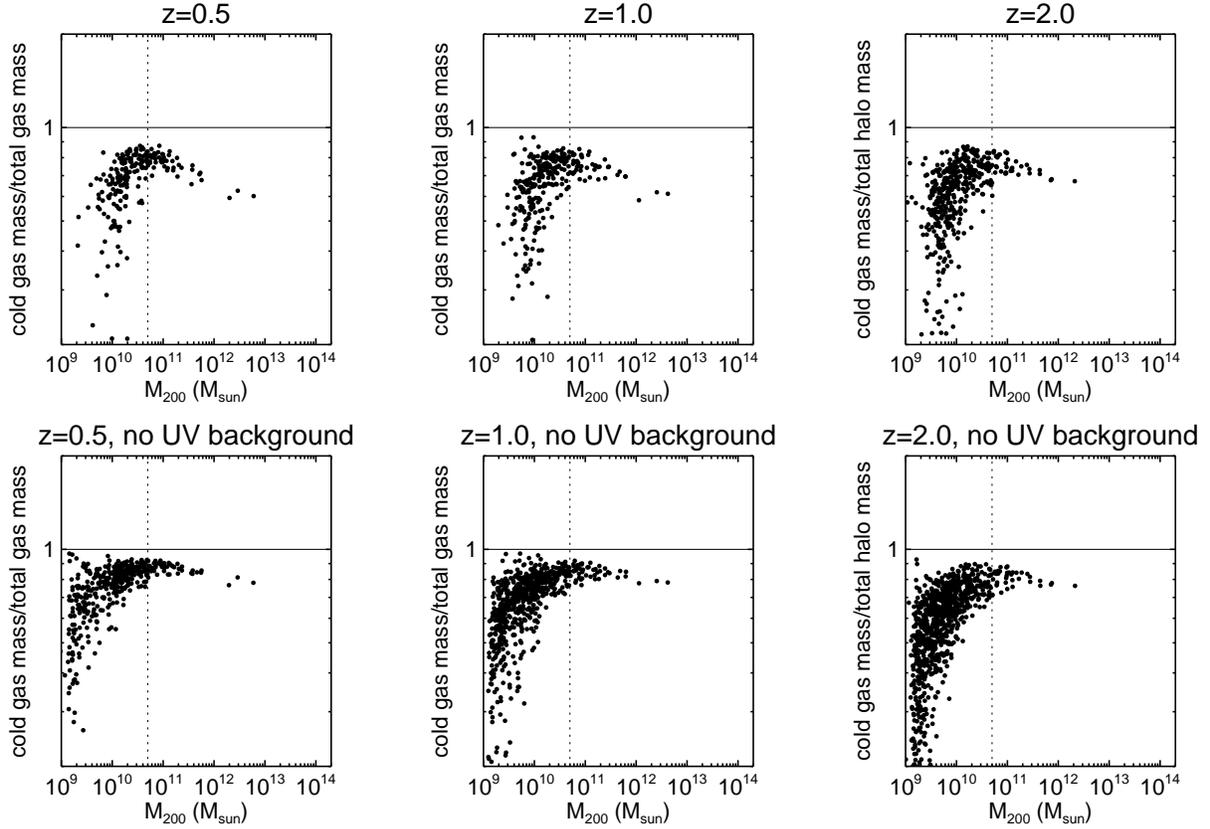}}
\caption{Ratio of cooled and total gas mass versus halo mass for
  all progenitors of all halos with $M_{200}>M_{\rm min\ halo}$, at
  three redshifts ($z=0.5$, $1$ and $2$), for the small 9 {\hMpc} box.
  The vertical dashed line denotes $M_{\rm min\ prog} =
  5\times10^{10}\ {\rm M}_\odot$, the horizontal thin line denotes
  unity.  The lower panels refer to the same simulation run without
  the UV background.}
\label{fig:minmass}
\end{figure*}

\subsection{Runs}
\label{section:runs}

Initial conditions were created using the public N-GenIC code of
V. Springel\footnote{
http://www.mpa-garching.mpg.de/gadget/right.html
},
which uses the Zel'dovich approximation to displace particles from a
regular cubic grid.  We assumed a $\Lambda$CDM cosmological model with
$\Omega_0=0.24$, $\Omega_\Lambda=0.76$, $\Omega_b=0.04$, $H_0=72$ km
s$^{-1}$ Mpc${-1}$, $\sigma_8=0.8$, $n_s=0.96$.  We generated initial
conditions for a box of 36 {\hMpc} (50 Mpc) sampled with $512^3$ DM
and $512^3$ gas particles.  The initial redshift was set equal to 199,
where the r.m.s. particle displacement for the large box is 30 per
cent of the grid spacing.  Particle masses are $1.93\times10^7$
{\hMsun} for DM and $3.86\times10^6$ {\hMsun} for gas.  With this
choice, a $10^{12}$ {\hMsun} halo is resolved with $\sim$50000 DM
particles.  The assumed value for the gravitational softening is 1.5
comoving {\hkpc}.  These values were found in \cite{Viola08} to be
adequate to fully resolve the cooling region of a $10^{12}$ {\hMsun}
halo.

We run this simulation in four configurations. (i) PUREDM: a pure
N-body simulation, where all gas particles are treated as
collisionless particles.  This was used to create merger trees for the
SAMs.  (ii) NOCOOL: a simulation with gas hydrodynamics but without
radiative cooling and UV heating.  (iii) COOL: a simulation with
radiative cooling, UV heating and ``star formation''. (iv) COOLZ2: a
simulation with the same physics as COOL, using a snapshot of the
NOCOOL simulation at $z=2.089$ as initial conditions.  In this paper
we present results based on the PUREDM, COOL and COOLZ2 simulations;
the NOCOOL simulation will be used in future papers.

To have a proper time sampling of the merger trees \citep{Benson12},
halo and substructure-finding were performed 128 times, uniformly
spaced in the logarithm of the scale factor from $a=0.05$ to $a=1$.
In this way, a halo is sampled roughly two times per dynamical time.
This ``post-processing'' was performed on-the-fly and included a
standard friends-of-friends (FoF) halo finder algorithm as well as the
substructure finder algorithm SUBFIND \citep{Springel01}.  To limit
storage requirements, snapshots were saved only 64 times, At $z=0$,
there are 16 halos more massive than $10^{13}$ {\hMsun} in the PUREDM
simulation, 159 more massive than $10^{12}$ {\hMsun} and 302 more
massive than $M_{\rm min\ halo}=4.6\times10^{11}$ \hMsun. This mass
roughly corresponds to 20000 DM particles, and we set it as a lower
limit for the mass of the main progenitor.  The number of halos that
we find is sufficient to span two orders of magnitude in halo mass and
perform averages over many halos.

Halos in the COOL and COOLZ2 simulations were matched to those of
PUREDM simulation by checking positions and masses.  A few ambiguous
cases were found whenever pairs of nearby halos were classified as
merged in one simulation and separated in the other; these halos were
removed from the catalog.

\subsection{Merger trees and halo masses}
\label{section:trees}

The construction of merger trees is affected by a number of issues
described, for instance, in \cite{Fakhouri08} or \cite{Tweed09}
\citep[see also][]{Srisawat13}.  Of the three SAMs used here, the
{\del} and {\gal} models run by default on merger trees based on dark
matter substructures, while {\mor} runs only on FoF-based trees. In
Paper I, all models were run on FoF-based trees, so as to carry out a
fair comparison; this choice was tested and discussed there. We adopt
the same approach in this paper.

We constructed FoF-based merger trees by directly matching pairs of
progenitor/descendant FoF halos in consecutive outputs.  We set a
minimum mass of 100 particles for the smallest progenitor considered.
Halo pairs were matched when they overlapped by more than 50 per cent
of particles, with respect to either the progenitor or the
descendant. When substructures are neglected, the main issue is the
splitting of halos, i.e. the case of FoF halos having two or more
descendants.  This can be due to either (i) artificial bridging of two
separated halos in one snapshot, which is a byproduct of the FoF
algorithm; or (ii) highly eccentric orbits of substructures that lead
them temporarily out of their main halo.  Assuming that a halo $A_i$
at the output {\it i} splits into halos $A_{i+1}$ and $B_{i+1}$ at
$i+1$ ($A_{i+1}$ being the largest descendant), we recognize $B_{i+1}$
as an artificially bridged halo (case i) if it verifies these
relatively conservative criteria: (1) both $A_{i+1}$ and $B_{i+1}$
overlap by at least 75 per cent (of their particles) with two separate
halos $A_{i-1}$ and $B_{i-1}$ in output $i-1$; (2) the mass of
$B_{i+1}$ is at least 1 per cent of that of $A_{i+1}$; (3) $B_{i+1}$
contains at least 300 particles; (4) the mass of $B_{i+1}$ is not
lower than 90 per cent of that of $B_{i-1}$.  Splitting is then solved
by creating a new $B_i$ halo that takes part of the mass of $A_i$,
descends from $B_{i-1}$ and is progenitor of $B_{i+1}$.  Because
bridged halos are likely to merge in the future, the risk of adopting
too conservative criteria is to anticipate the merging time in some
cases; we think that this is more acceptable than attempting to
separate halos that have truly merged.  When these criteria are not
fulfilled, the halos are considered to belong to case (ii), that is
solved by absorbing the split (smaller) halo $B_{i+1}$ back into the
main halo $A_{i+1}$, then iterating the procedure for all descendants
of $B_{i+1}$.

Idealized merger trees obey two further conditions: halo masses should
never decrease with time and, after each merger, the mass of the
remnant should be larger than that of the merging halos.  Merger trees
extracted from simulations do not follow these rules.  In SAMs, the
amount of baryons associated with a halo is computed as the halo mass
multiplied by the universal baryon fraction, so if the former
decreases the SAM would lose baryons. In reality, a decrease in halo
mass is typically due to a readjustment of the halo profile after a
merger, so it is realistic to assume, as we did in Paper I, that
whenever halo mass decreases the baryonic mass available to the halo
remains constant.  The amount of baryons is thus determined as the
cosmic baryon fraction times the largest value of the halo mass
reached along the main progenitor branch, including the mass of
merging progenitors.

As for the halo mass, we use {\mtwoh}, computed as the
mass within a sphere (of radius $r_{200}$) such that its overdensity
is 200 times the {\em critical} density of the Universe,
$\rho_c=3H^2/8\pi G$.  This quantity is computed by the SubFind code,
using as center of the sphere the most bound particle of the main
substructure of the FoF halo.  Fig.~\ref{fig:mah0} shows the mass
accretion history of the main progenitor of the most massive halo in
the PUREDM simulation, hereafter named halo 0.  At $z=0$ this halo has
$M_{200}=6.49 \times 10^{13}\ {\rm M}_\odot$.  In this figure the black
continuous and the blue dotted lines correspond to {\mtwoh} and {\mfof} of the
main progenitor in the PUREDM simulation.  The vertical black dashes mark the
times corresponding to major mergers, i.e.  mergers of halos with mass ratios
larger than 1:10.  The length of the dash is proportional to the mass ratio.
The two masses grow similarly in time, with the difference that, at major
mergers, {\mtwoh} responds more slowly than {\mfof} to the increase of mass.  By
visual inspection, we verified that the FoF algorithm merges two halos when they
start to overlap, but the mass of the smaller progenitor significantly overlaps
with the $r_{200}$ sphere only towards the end of the relaxation process, which
lasts $\sim2-3$ halo dynamical times.  The difference in the two growth curves
is a reflection of the fact that there is no obvious definition of ``mass'' for
a halo undergoing a major merger.

Fig.~\ref{fig:mah0} also shows the mass accretion histories (defined
via \mtwoh) of the same halo in the COOL (green dashed line) and
COOLZ2 (red dot-dashed line) simulations. Because cooling concentrates
baryonic matter in the halo center, the {\mtwoh} masses in the cooling
runs are slightly larger -- $\la 10$ per cent -- than in the PUREDM
run.  Apart from this, the merger trees are very similar.  This degree
of similarity is observed in most halos, with a few exceptions in
cases where, at high redshift ($z>2$), two merging halos have very
similar masses and the most massive progenitor is not the same in the
two simulations.  These results confirm that using merger trees from a
collisionless simulation induces only small differences in model
results.

\begin{figure*}
\centering{\includegraphics[width=.95\textwidth]{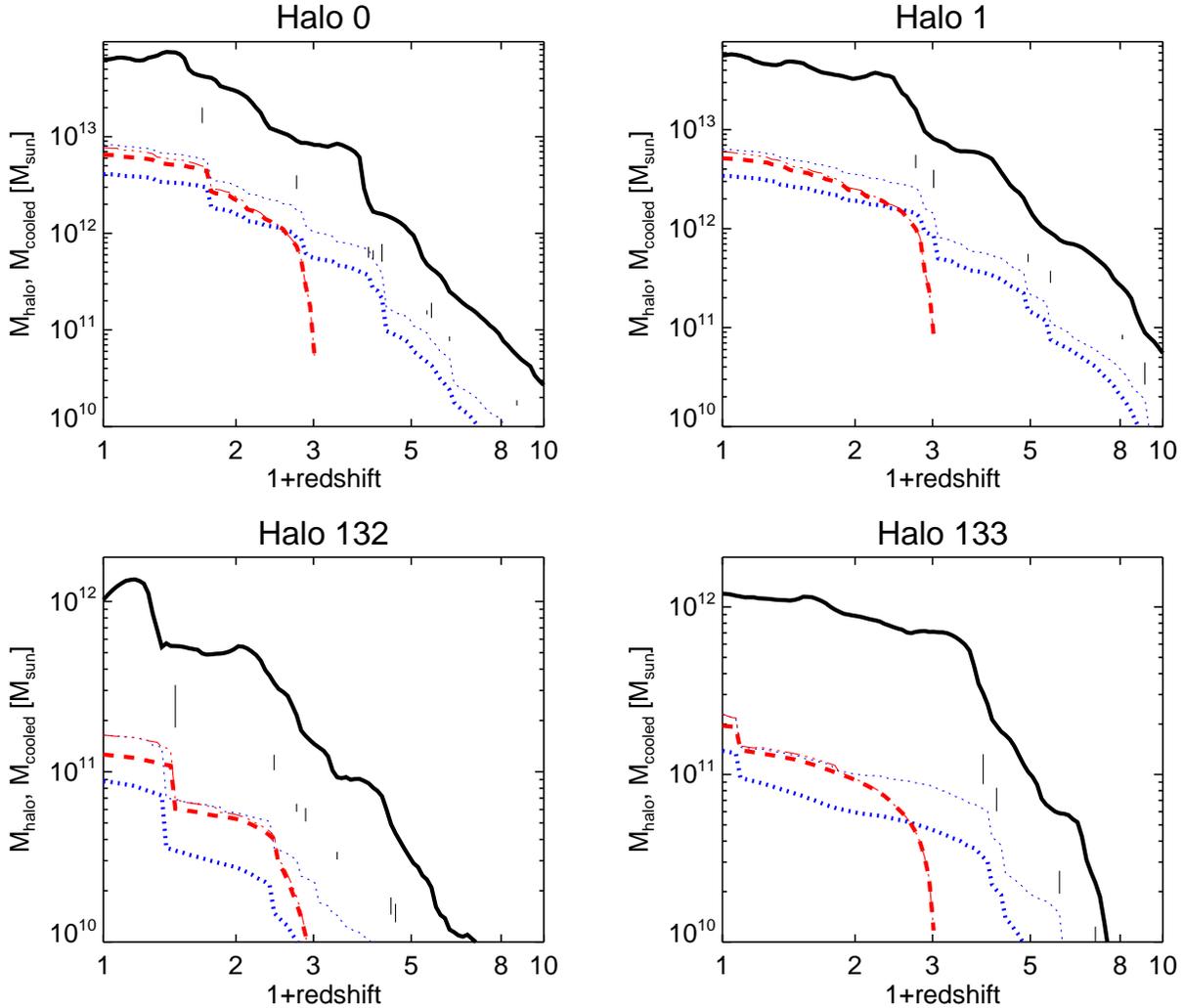}}
\caption{DM and cooled mass accretion histories of the main
  progenitors of the four example halos.  Black lines give DM masses,
  with thin vertical dashes marking the position of major mergers.
  For the colored lines, thick lines give the mass of cooled gas
  computed by summing the cooling rates over all progenitors more
  massive than $M_{\rm min\ prog}$ while thin lines give the total
  cooled mass.  Blue dotted lines refer to the COOL simulation and red
  dashed lines refer to the COOLZ2 simulation.}
\label{fig:mah_massive}
\end{figure*}

\subsection{Computation of cooled masses}
\label{section:cooled}

The cooled mass of a halo can be simply computed as the total mass of
``star'' particles.  These particles cool in the deepest regions of
potential wells, but then, as noticed in, e.g., \cite{Saro10}, they
are subject to tidal forces, if not to two-body heating, and scatter
away from the substructure in which they were born to form a ``diffuse
stellar component''.  Some of these stars can scatter beyond
$r_{200}$, thus making the computation of cooled mass more
complicated.  The cooling rate onto a halo can then be computed by
taking the difference of cooled masses between two consecutive
outputs; in this case it is necessary to subtract the contribution
from gas that cooled in other progenitors and then entered the halo
via merging.  With post-processing performed on time intervals smaller
than the halo dynamical time, we resort to an alternative computation
of cooled masses, which avoids the difficulties mentioned above and
allows us to easily subtract out the contribution from insufficiently
resolved halos.

{\sc GADGET} assigns to each cooled particle (``star'') an ``age'',
equal to the scale factor at ``star formation''.  Because particles
retain their ID number, we identify for each cooled particle the FoF
halo it belonged to at the last output where it was still a gas
particle.  We find that only 0.3 and 0.4 per cent of particles in the
COOL and COOLZ2 simulations did not belong to any FoF halo at the last
output before cooling.  The cooling rate of a halo at a given output
is then computed as the mass of all cooled particles that were gas
particles in the previous output and belonged to the halo main
progenitor, divided by the time interval between the two outputs.
Cold masses are computed by summing cooling rates over all progenitors
along the merger tree.

This approach allows us to subtract out the contribution coming from
poorly resolved halos.  In previous work
\citep[e.g.][]{Benson01,Yoshida02,Helly03}, based on simulations at
much poorer resolution, this limit was often estimated by requiring
that the halo has at least as many gas particles as the number of
neighbors used to compute the density (64 in our case).  For our
simulation, this would give $\sim10^9$ \hMsun.  This is a very
optimistic estimate, because cooling takes place in the inner region
of the halo, whose size is of order of the scale radius.  
To quantify the mass scale above which the cooling region is well
resolved in our simulations, we use the 9 {\hMpc} box already
discussed in Section~\ref{section:code}, run with the same setting as
the COOL simulation (identical conclusions are reached using the full
COOL and COOLZ2 simulation).  We show in the upper panels of
Fig.~\ref{fig:minmass}, for all progenitors of all halos more massive
than $M_{\rm min\ halo}$ and for three redshifts ($z=0.5$, $1$ and
$2$), the ratio between cooled mass and total gas mass in the halo.
At small masses, below roughly $M_{\rm min\ prog}=5\times10^{10}$
\msun, cooled fractions begin to drop to low values.  This
behavior is specific to the simulation and is not observed in SAM
predictions.  This drop is in part an effect of heating due to the UV
background.  To subtract out this effect, the lower panels report the
same quantities for a simulation run {\em without} a UV background.
In this case cooled masses change significantly below $10^{10}$
{\msun}, but the change of trend with halo mass at $M_{\rm min\ prog}$
is still present and has to be ascribed to limited numerical resolution.

A halo of mass $M_{\rm
  min\ prog}$ is sampled by $\sim$10000 DM particles, and at $z=0$ it
has a virial radius of $\sim75$ kpc; for a concentration of $\sim10$,
the scale radius is $\sim7.5$ kpc, $\sim5$ times the gravitational
softening.  This is a good, conservative, lower limit for the
mass of the smallest halo where cooling is properly resolved in this
simulation.

From now on, unless otherwise specified, cooled masses will be
computed by summing cooling rates on all progenitors more massive than
$M_{\rm min\ prog}$.  In Fig.~\ref{fig:mah_massive} we show the cooled
mass found in the main progenitor of four example halos that will be
used in the following analysis.  Two of these are the most massive
halos in the simulation, halo 0 and halo 1, with the former suffering
a major merger below $z=1$ and the second having a more quiet merger
history at late times. The other two halos have masses
$\sim10^{12}\ {\rm M}_\odot$, with halo 132 suffering a recent major
merger and halo 133 having a quiet mass accretion history.  For each
halo, we show results from the COOL (dotted blue lines) and COOLZ2
(red dashed lines) simulations. Thick lines give cooled masses
computed as described above while thin lines give the total mass of
cooled particles in the halos; accordingly, the differences between
the thin and thick lines give the contributions of poorly resolved
halos.  This is very significant in all cases for the COOL simulation,
amounting to roughly a factor of two, while it is much more limited
for in the COOLZ2 simulation. It is worth reminding that this increase
of cooled mass is contributed both by cooling and by mergers.

This figure demonstrates the ability of COOLZ2 to suppress
over-cooling in poorly resolved progenitors. In the COOL simulation
the fraction of cold mass that has cooled in well-resolved progenitor
halos is roughly constant with redshift (and of order of $\sim1/2$),
because small halos continually bring cold gas into the main
progenitor through mergers.  The COOLZ2 simulation produces total
cooled masses (the thin lines) that by $z=0$ have converged to those
of the COOL simulation.  This is expected, because the fraction of the
residual hot gas will be roughly determined by the density at which
cooling times become equal to the Hubble time.  But in COOLZ2 most gas
cools in haloes with $M_{200} > M_{\rm min\ prog}$, since high-$z$
overcooling at high redshift is removed and the hot gas density at
$z\sim2$ is higher than in COOL.  
We thus conclude that in this simulation the
numerical description of cooling is less affected by poor resolution
than in the COOL run.

The computation of hot masses is affected by an uncertainty related to
the definition of halo mass: models use {\mtwoh} to compute the total
amount of available baryons, so the output of the post-processing
gives the mass of hot gas within the FoF halo.  Then, a comparison of
models with simulations is affected by the scatter between {\mtwoh}
and {\mfof}.  This problem affects the computation of cooled mass
described above to a much smaller extent, since particles that suffer
runaway cooling are located well within $r_{200}$.  We have verified
that the behavior of hot masses is entirely predictable from that of
cooled masses, so for sake of simplicity we will restrict our analysis
to cold masses and cooling rates.

\section{Semi-analytic models}
\label{section:sam}

We compare the results of the simulations described above with
predictions from the three stripped-down SAMs {\gal}, {\mor} and
{\del} as defined in the Introduction.  All models have been adapted
to run on the FoF merger trees of the simulation.  In all cases, the
only baryonic processes implemented are shock heating, cooling and
infall of gas onto the central galaxy.  Cooling times are computed
using the \cite{SD93} cooling functions assuming zero metallicity.
Merger times are set to zero so that galaxy mergers immediately follow
the merging of DM halos.  Predictions for the COOLZ2 simulation are
obtained by switching on gas cooling only after $z=2.089$.  As for the
cooling rates, we compute them as the cold mass accumulated on the
central galaxy between two snapshots, divided by the time interval
between them.  We verified that these cooling rates are very similar
to the instantaneous cooling rates evaluated at the time of the later
snapshot.  The cooling and infall models embedded in these SAMs are
described at length in the original papers and in Paper I, so we give
only a short summary of their main properties here.

\subsection{The \emph{delucia} model}

The rate of gas cooling is computed following the model originally
proposed by \citet{White91}, with the specific implementation
following that of \citet{Springel01}.  The hot gas within dark matter
halos is assumed to follow an isothermal profile:

\begin{equation}
\rho_{\rm g}(r) = \frac{M_{\rm hot}}{4\pi r_{200} r^2}.
\end{equation}

\noindent
At each output, the total amount of hot gas available for cooling in
each halo is estimated as follows:

\begin{equation}
  M_{\rm hot} = f_{\rm b}M_{200}-M_{\rm cold}
\label{eq:mhotcompute}
\end{equation}

\noindent
where $f_{\rm b}$ is the universal baryon fraction and $M_{\rm cold}$
is the cold mass associated to the halo.  Eq.~\ref{eq:mhotcompute} can
provide, in a few cases, a negative number (this occurs typically
after important halo mergers).  In this case, the amount of hot gas is
set to zero, and no cooling is allowed in the remnant halo.

The equations for the evolution of gas are solved using $20$
time-steps between each pair of simulation snapshots.  The cooling
time is defined as the ratio of gas specific thermal energy and
cooling rate per unit volume:

\begin{equation}
  t_{\rm cool}(r) = \frac{3}{2} \frac{kT\rho_{\rm g}(r)}{\bar{\mu}m_{\rm
  p}n_{\rm e}^2(r)\Lambda(T,Z)}
\label{eq:tcool}
\end{equation}

\noindent
In the above equation, $\bar{\mu}m_{\rm p}$ is the mean particle mass,
$n_{\rm e}(r)$ is the electron density, $k$ is the Boltzmann constant,
and $\Lambda(T,Z)$ represents the cooling function for a zero
metallicity gas.  The virial temperature of the halo is computed as:

\begin{equation}
T_{\rm vir} = \frac{1}{2} \frac{\mu m_{\rm H}}{k} V_{\rm vir}^2\,\,\,\,\,\,\,{\rm
  or}\,\,\, T_{\rm vir} = 35.9(V_{\rm vir}/{\rm km}\,{\rm
  s}^{-1})^2\,\,\,{\rm K}
\label{eq:Tvir}
\end{equation}

\noindent
where $m_{\rm H}$ is the mass of the hydrogen atom, and $\mu$ is the mean
molecular mass. 

A `cooling radius' is then computed as the radius at which the local
cooling time is equal to the halo dynamical time.  If the cooling
radius is smaller than the virial radius of the halo under
consideration, the gas is assumed to cool quasi-statically, and the
cooling rate is modeled by a simple inflow equation:

\begin{equation}
\frac{{\rm d}M_{\rm cool}}{{\rm d}t} = 4\pi\rho_{\rm g}(r_{\rm cool})r_{\rm
  cool}^2\frac{{\rm d}r_{\rm cool}}{{\rm d}t} ,  \;\hbox{ if } r_{\rm
    cool} < r_{\rm vir}.
\label{eq:coolingrate}
\end{equation}

\noindent
At early times, and for low-mass halos, the formal cooling radius can
be much larger than the virial radius. In this infall-dominated
regime, the infalling gas is assumed not to reach hydrostatic
equilibrium and the accretion rate on the galaxy is determined by the
halo dynamical time:

\begin{equation}
\dot{M}_{\rm cool} = M_{\rm hot}/\tau_{\rm dyn}  \; \hbox{ if }
  r_{\rm cool} \ge r_{\rm vir}\,,
\label{eq:infallrate}
\end{equation}

\noindent
where $\tau_{\rm dyn}=R_{\rm vir}/V_{\rm vir}$.

\subsection{The \emph{galacticus} model}

The highly modular {\gal} code has been configured so as to reproduce
the behavior of the Durham model of \cite{Cole00}, \cite{Benson03}
and \cite{Bower06}.

The hot gaseous component in dark matter halos is assumed to have a
density profile described by the $\beta$-model:

\begin{equation}
\rho_{\rm g}(r) = \frac{\rho_0}{[1+(r/r_{\rm core})^2)]^{3\beta/2}}
\end{equation}

\noindent
where $\rho_0$ is the density at the center of the halo, $r_{\rm
  core}$ is the radius of the core, and $\beta$ is a parameter that
sets the slope of the profile at $r\gg r_{\rm core}$.  The model
assumes $\beta=2/3$; the standard choice of core radius is $r_{\rm
  core} = 0.07 \cdot r_{200}$.  The temperature profile of the gas is
assumed to be isothermal at the virial temperature (Eq.~\ref{eq:Tvir})

Evolution is integrated between each pair of simulation snapshots
using an adaptive timestep integrator set to maintain a tolerance of 1
part in $10^3$ in all evolved quantities. Halo masses are linearly
interpolated between their values at successive snapshots when needed
as input to the integrator.

When the mass of a halo is increasing with time, $\dot{M}_{200} > 0$,
the hot halo gains mass at a rate
\begin{equation}
\dot{M}_{\rm hot} = f_{\rm b} \dot{M}_{200}.
\end{equation}
In cases where the the halo is decreasing in mass with time,
$\dot{M}_{200}<0$, this mass loss is instead accumulated to a
quantity, $M_{\rm loss}$:
\begin{equation}
\dot{M}_{\rm loss} = - f_{\rm b} \dot{M}_{200},
\end{equation}
while $M_{\rm hot}$ remains constant (or decreases if cooling is
occurring). Once the halo begins to grow once more, the accretion rate
onto the halo becomes
\begin{eqnarray}
\dot{M}_{\rm hot}  &=& \left( f_{\rm b} + {M_{\rm loss}\over M_{200}} \right) \dot{M}_{200}, \\
\dot{M}_{\rm loss} &=& \left(- {M_{\rm loss}\over M_{200}} \right) \dot{M}_{200},
\end{eqnarray}
such that the mass of hot gas does not begin to significantly grow
until this halo exceeds the previous maximum value of $M_{200}$.

The cooling time is defined using Eq.~\ref{eq:tcool}.  The cooling
radius is computed by equating the radial-dependent cooling time with
the dynamical time of the halo. The cooling rate is then given by
Eq.~\ref{eq:coolingrate} if $r_{\rm cool} < r_{\rm vir}$ or by
Eq.~\ref{eq:infallrate} if $r_{\rm cool} \ge r_{\rm vir}$.  The mass
of the hot halo is decreased at this rate, and $\dot{M}_{\rm cool}$ is
integrated between each pair of simulation snapshots to find the mass
of gas that flows onto the central galaxy.

\subsection{The \emph{morgana} model}
\label{sec:morgana}

The cooling model implemented in {\mor} is described in
\cite{Monaco07} and \cite{Viola08}.  The hot halo phase is assumed to
be in hydrostatic equilibrium in a NFW halo, to fill the volume
between the cooling radius and the virial radius of the halo, and to
be subject to a polytropic equation of state with index $\gamma_p =
1.15$.  Under these assumptions, one obtains:

\begin{equation}
\rho_{\rm g}(r) = \rho_{g0} \left(1-a\left(1-\frac{\ln (1+c_{\rm nfw}x)}{c_{\rm
    nfw}}\right)\right)^{1/(\gamma_p -1)}
\label{eq:rho_mor}
\end{equation}
\begin{equation}
T_{\rm g}(r) = T_{g0}\left(1-a\left(1-\frac{\ln (1+c_{\rm nfw}x)}{c_{\rm
    nfw}}\right)\right)
\label{eq:T_mor}
\end{equation}

\noindent
where $a$ is a function of halo concentration ($c_{\rm nfw}= r_{\rm
  halo}/r_s$), $x=r/r_s$ and $r_s$ is the halo scale radius.
$\rho_{g0}$ and $T_{g0}$ are the values of density and temperature
extrapolated to $r=0$, while $\eta(r) = T_{\rm g}(r)/T_{\rm vir}$ and
$\eta_0=\eta(0)$.

The cooling rate is computed by integrating the contribution to
cooling from all mass shells:

\begin{eqnarray}
\lefteqn{\frac{dM_{\rm cool}}{dt} = \frac{4 \pi r_s^3
    \rho_{g0}}{t_{{\rm cool,}0}} \times }\\
&&
\int_{r_{\rm cool}/r_s}^{c_{\rm nfw}}\left[1-a\left(1 -
  \frac{\ln(1+t)}{t}\right)\right]^{2/(\gamma_p-1)} t^2 dt
\end{eqnarray}

\noindent
where $t_{\rm cool,0}$ is computed using the central density $\rho_{g0}$ and
the temperature of the gas at $r_{\rm cool}$.

At each output, the mass of gas that falls onto the halo is computed
as:

\begin{equation}
 M_{\rm infall} = f_{\rm b} \cdot {\rm max} \left[ M_{200} - \left(\sum_i
   M_{200}^i + M^{\rm max}_{200} \right) , 0.0 \right]
\end{equation}

\noindent
where the sum extends over all progenitors of the halo other than the
main one, and $M^{\rm max}_{200}$ is the maximum value of the virial
mass along the main progenitor branch.  The evolution of the system is
followed with a Runge-Kutta integrator with adaptive time steps, and
the equilibrium configuration of the hot halo gas is re-computed at
each time-step.

By equating the mass cooled in a time interval $dt$ with the mass
contained in a shell $dr$, one obtains the evolution of the cooling radius,
which is treated as a dynamical variable in this model:

\begin{equation}
\frac{dr_{\rm cool}}{dt} = \frac{dM_{\rm cool}/dt}{4 \pi \rho_{\rm g}(r_{\rm
    cool}) r_{\rm cool}^2} - c_{\rm s}
\label{eq:coolingradius}
\end{equation}

\noindent
where $c_{\rm s}$ is the sound speed computed at $r_{\rm cool}$, and
is added to the right hand side of the above equation to allow the
`cooling hole' to close at the sound speed.

The calculation of cooling rate is started when a halo appears for the
first time, with $r_{\rm cool} = 0.001 \cdot r_s$, and is reset after
each halo major merger.  Cooling is then assumed to be quenched for
$n_{\rm quench}$ dynamical times after each merger, plus one cooling
time, to mimic the time needed by the merging system to relax into the
new configuration and start cooling again.  We use the standard value
of $n_{\rm quench}=0.3$.

\begin{figure*}
\centering{\includegraphics[width=.90\textwidth]{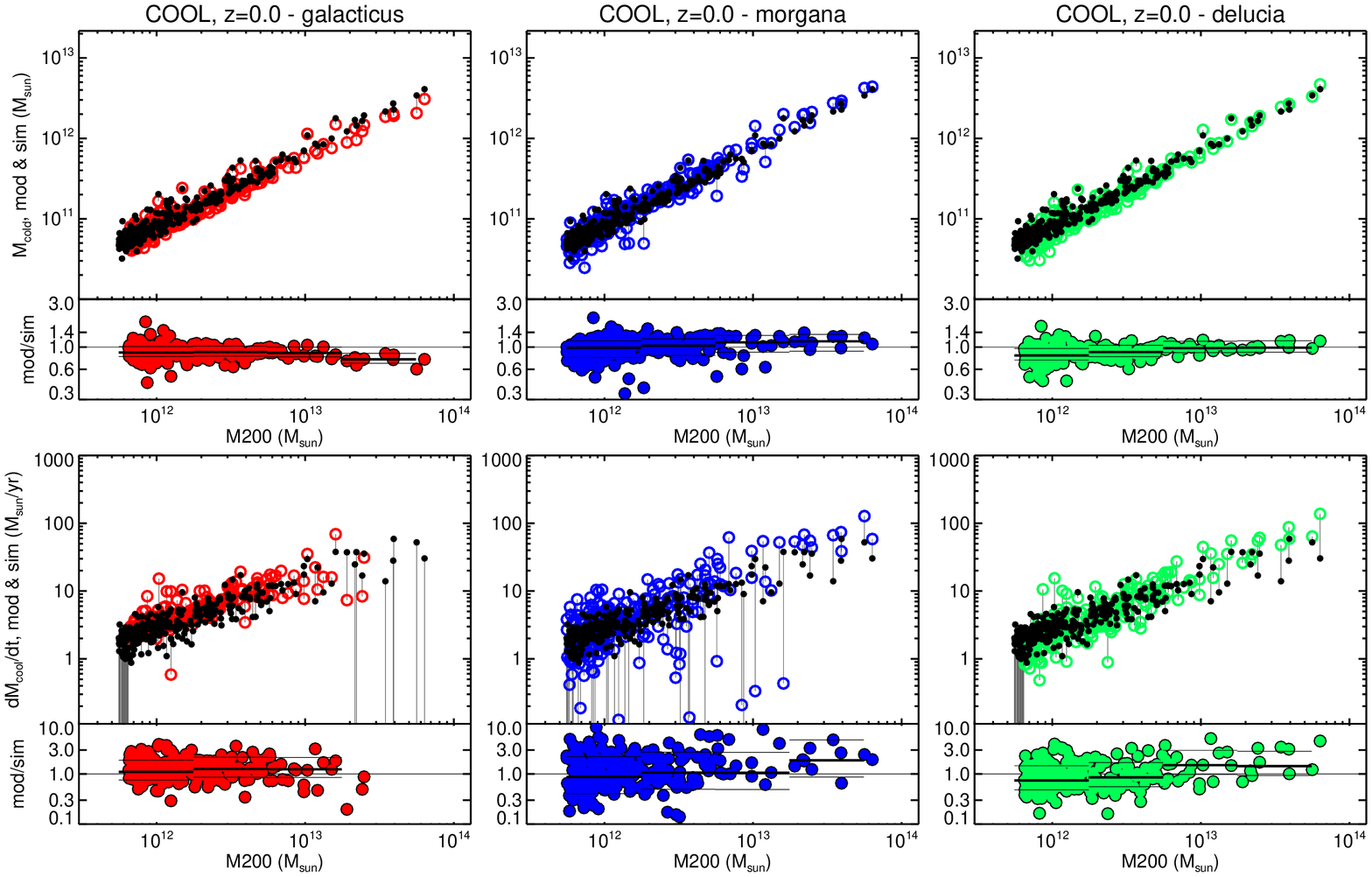}}
\centering{\includegraphics[width=.90\textwidth]{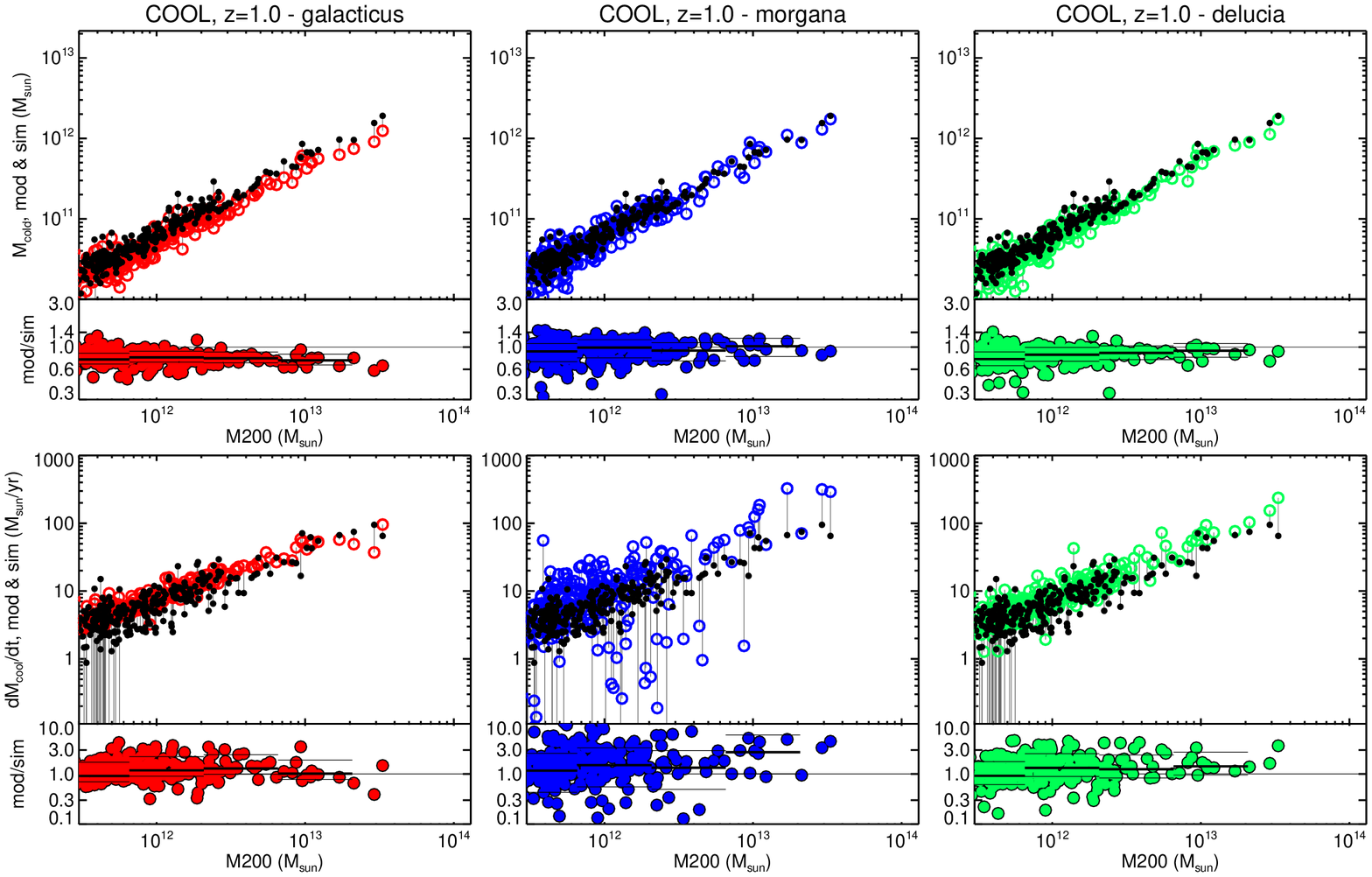}}
\caption{Cooled masses and cooling rates as predicted by models and
  simulations, for all the halos with mass higher than $M_{\rm
    min\ halo}$ and their main progenitors.  For each panel, the upper
  part gives, as function of $M_{200}$, the cooled mass or the cooling
  rate as found in the simulations (small black dots) and as predicted
  by the SAM (open colored circles).  Pairs of points corresponding
  to the same halo are connected by a thin black line; this allows to
  highlight the cases in which SAM predictions lie beyond the limits
  of the plot.  The lower part of each panel gives the ratio of the two quantities,
  again as a function of $M_{200}$, while the horizontal lines give
  medians and 16th and 84th percentiles in halo mass bins.  Each group
  of $3\times2$ panels reports cooled masses (upper three panels) and
  cooling rates (lower three panels) for the {\gal} (right panels, red
  points), {\mor} (mid panels, blue points) and {\del} (left panels,
  green points).  From the top, results are given at $z=0$ and $z=1$,
  while the second part of the figure gives the same quantities for
  $z=2$ and $z=3$.}
\label{fig:scatters_cool}
\end{figure*}

\begin{figure*}
\addtocounter{figure}{-1}
\centering{\includegraphics[width=.90\textwidth]{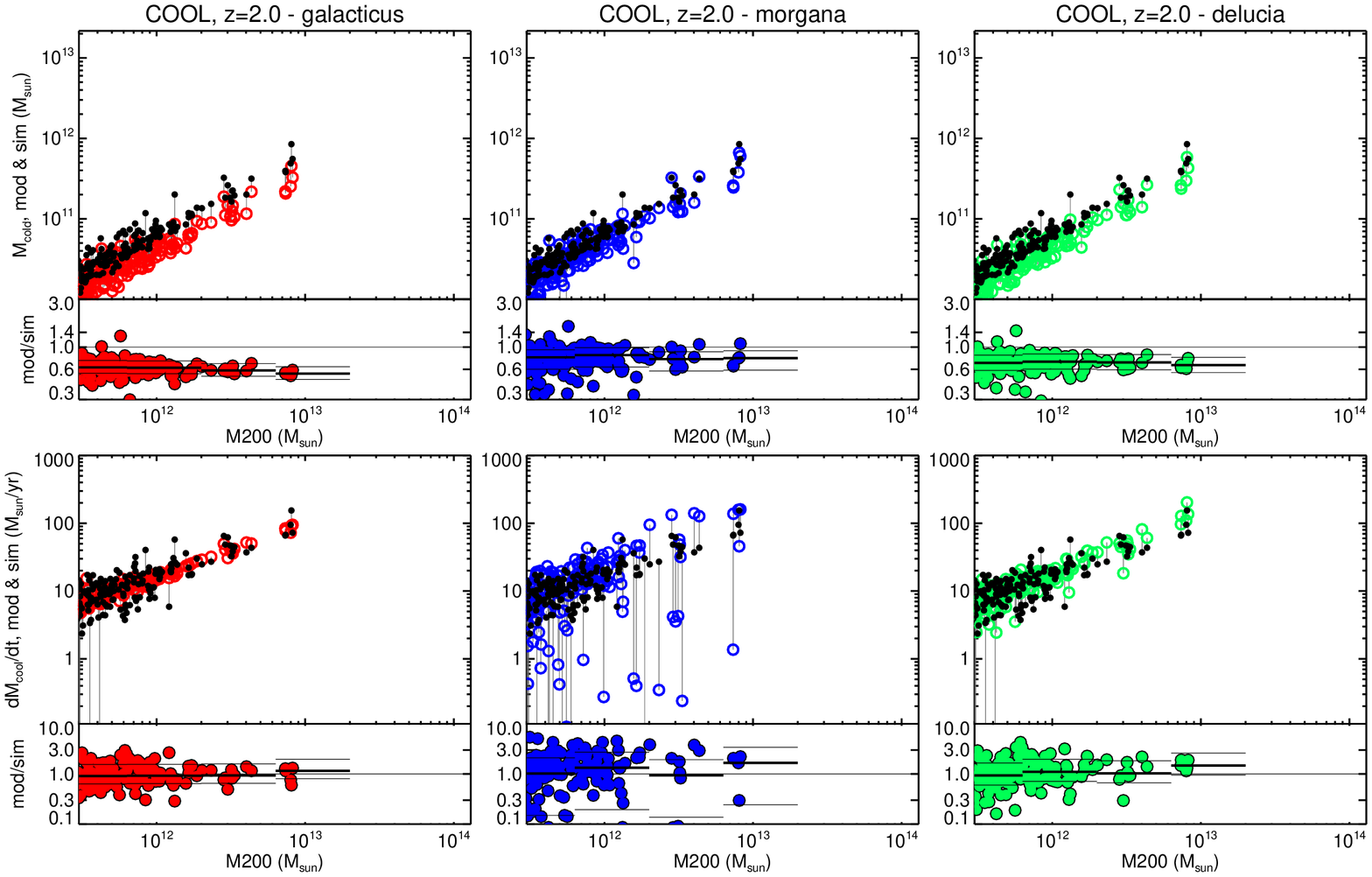}}
\centering{\includegraphics[width=.90\textwidth]{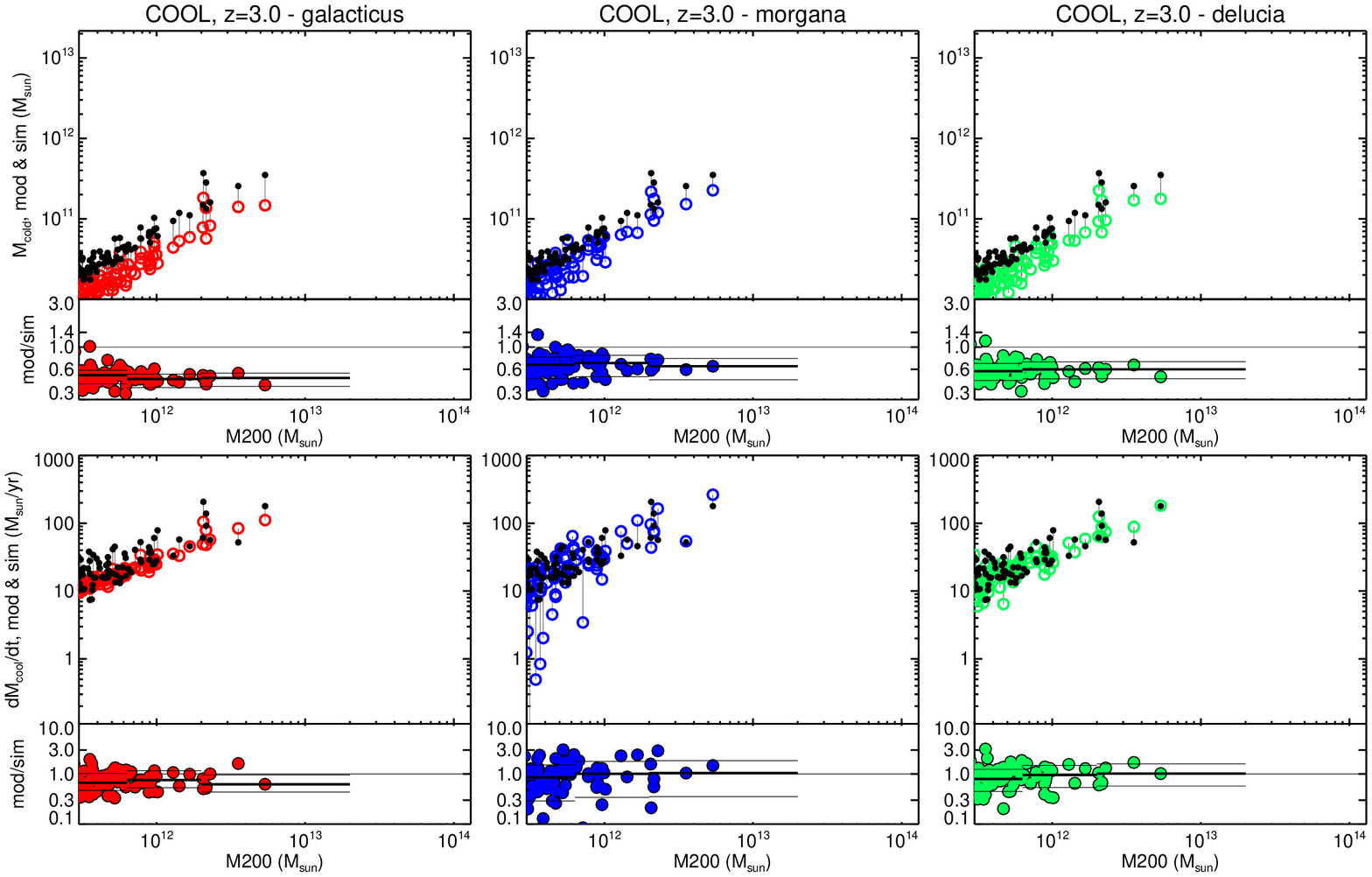}}
\caption{Continued.}
\end{figure*}

\begin{figure*}
\centering{\includegraphics[width=.90\textwidth]{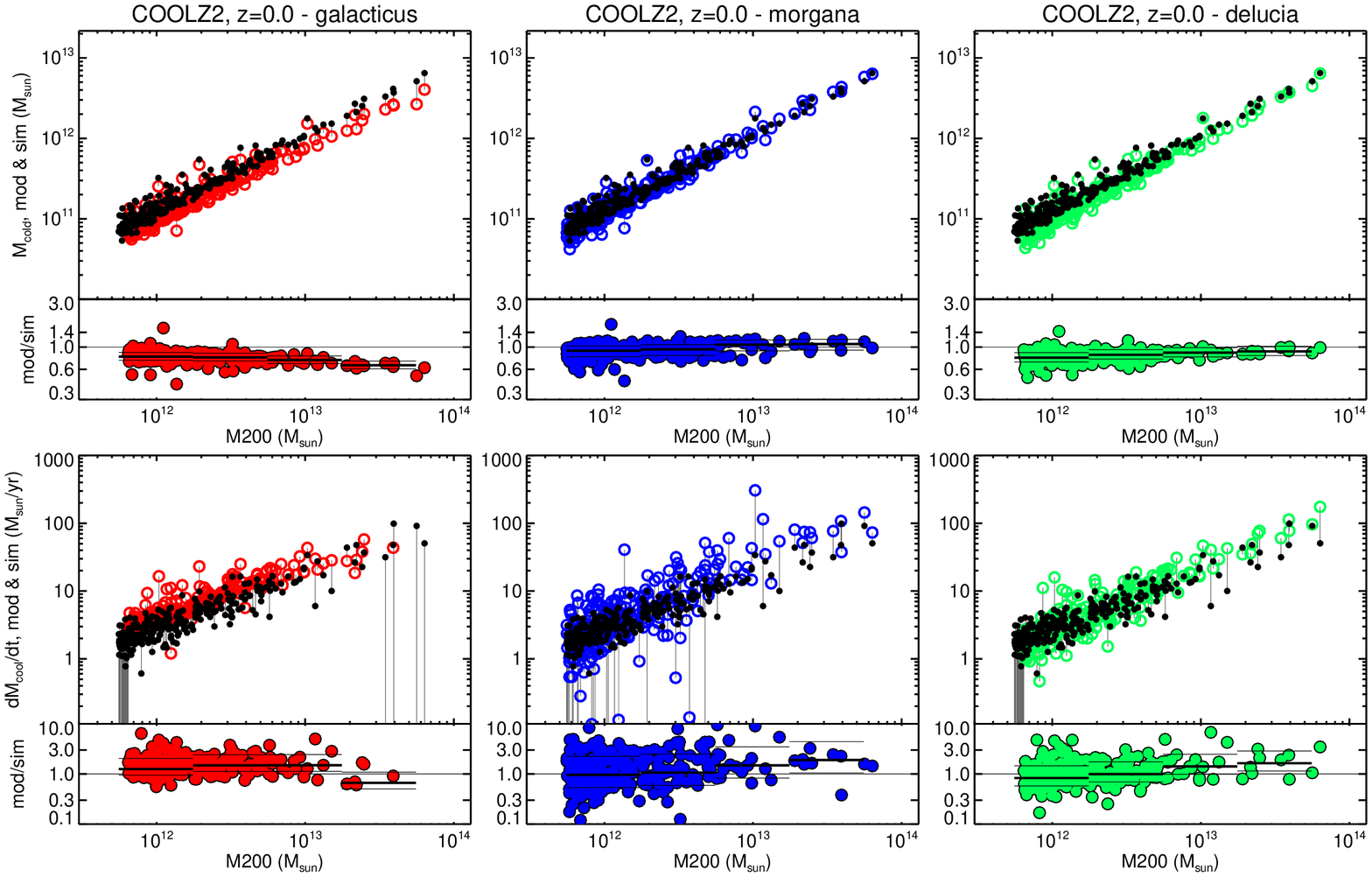}}
\centering{\includegraphics[width=.90\textwidth]{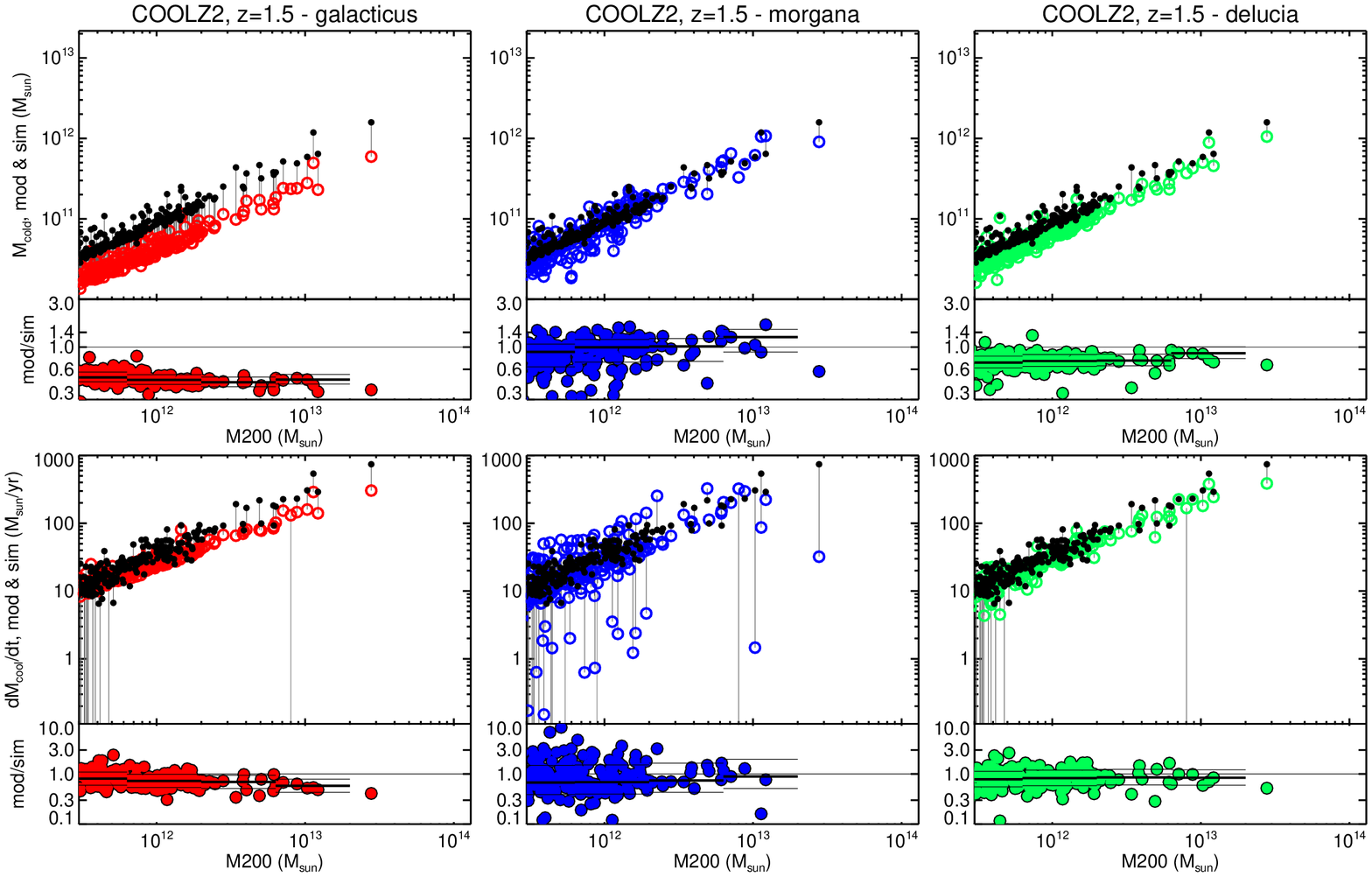}}
\caption{The same as Fig.~\ref{fig:scatters_cool}, for the COOLZ2 simulation. 
From the top, results are given at $z=0$ and $z=1.5$.}
\label{fig:scatters_z2}
\end{figure*}

\begin{figure*}
\centering{\includegraphics[width=.95\textwidth]{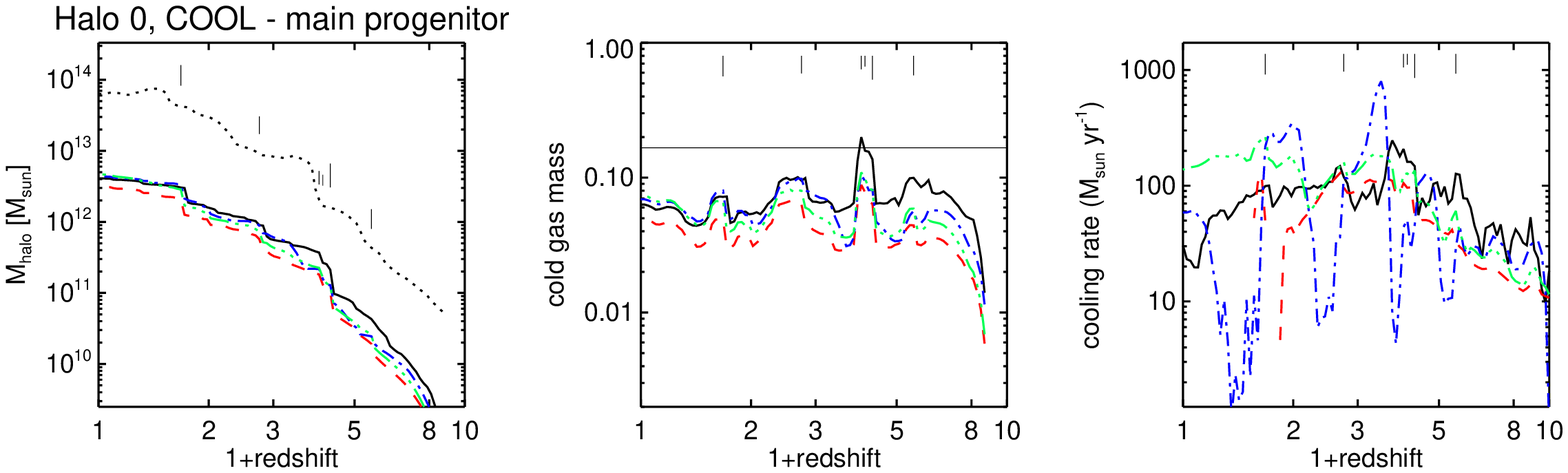}}
\centering{\includegraphics[width=.95\textwidth]{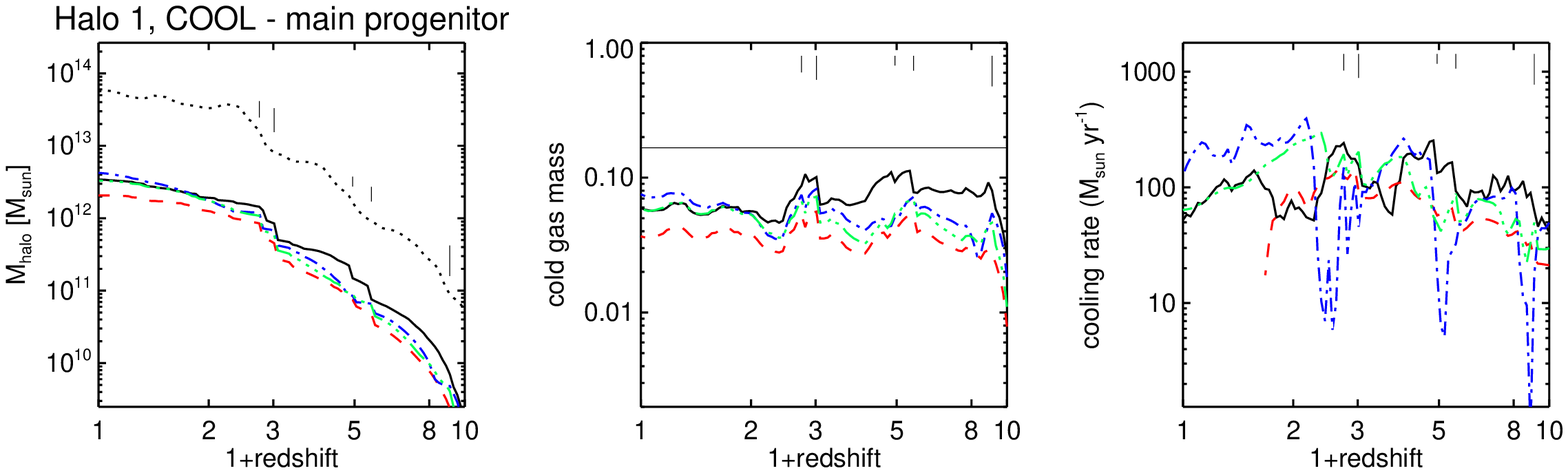}}
\centering{\includegraphics[width=.95\textwidth]{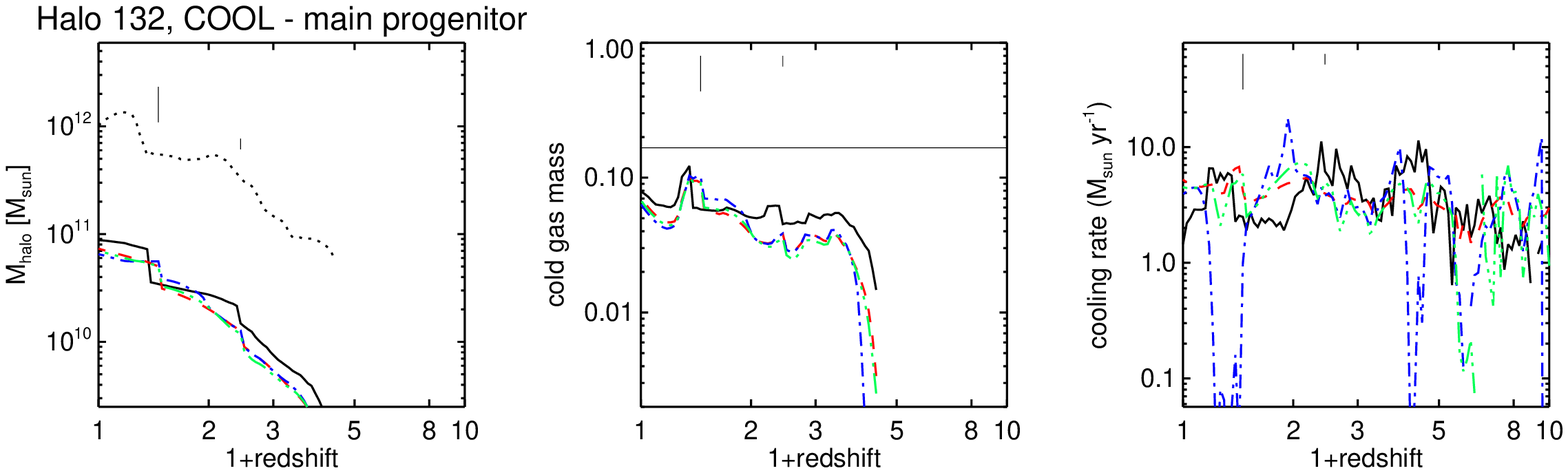}}
\centering{\includegraphics[width=.95\textwidth]{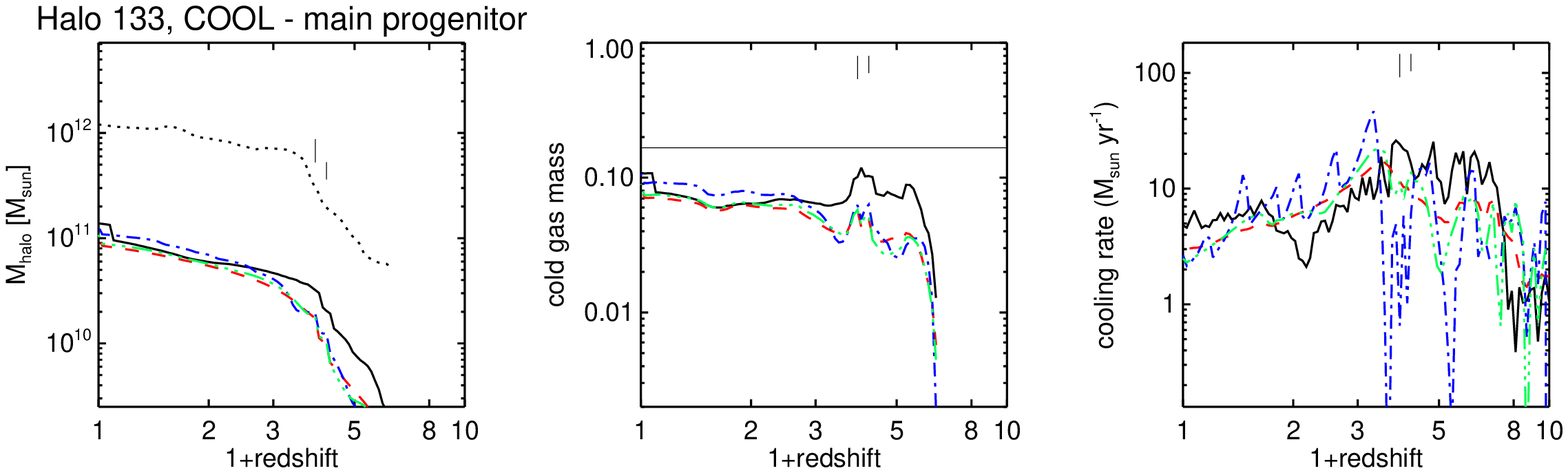}}
\caption{Evolution of cold gas mass, cold mass fraction and cooling rate of
  the main progenitor for the four example trees in the COOL
  simulation.  Black continuous lines report the quantities for the simulation,
  red dashed, blue dot-dashed and green triple dot-dashed 
lines give results for the {\gal}, {\mor}
  and {\del} models.  In the left panel, the dotted lines give
    the value of $M_{200}$.
In the mid panels the horizontal line denotes the
  universal baryon fraction. Thin vertical dashes mark major mergers.}
\label{fig:trees}
\end{figure*}

\begin{figure*}
\centering{\includegraphics[width=.95\textwidth]{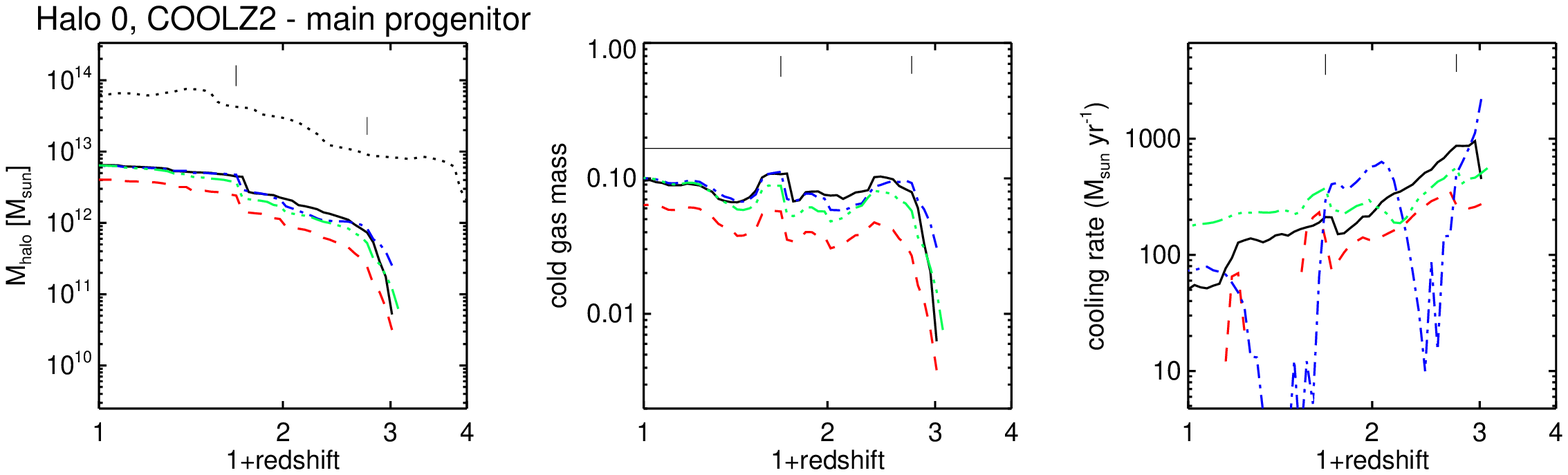}}
\centering{\includegraphics[width=.95\textwidth]{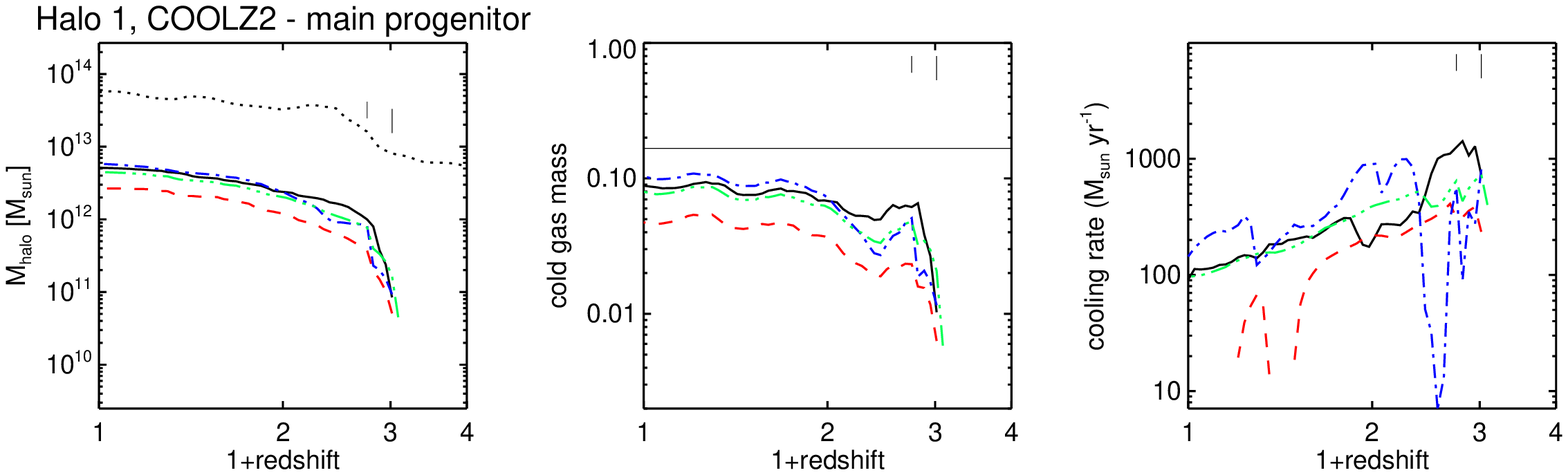}}
\centering{\includegraphics[width=.95\textwidth]{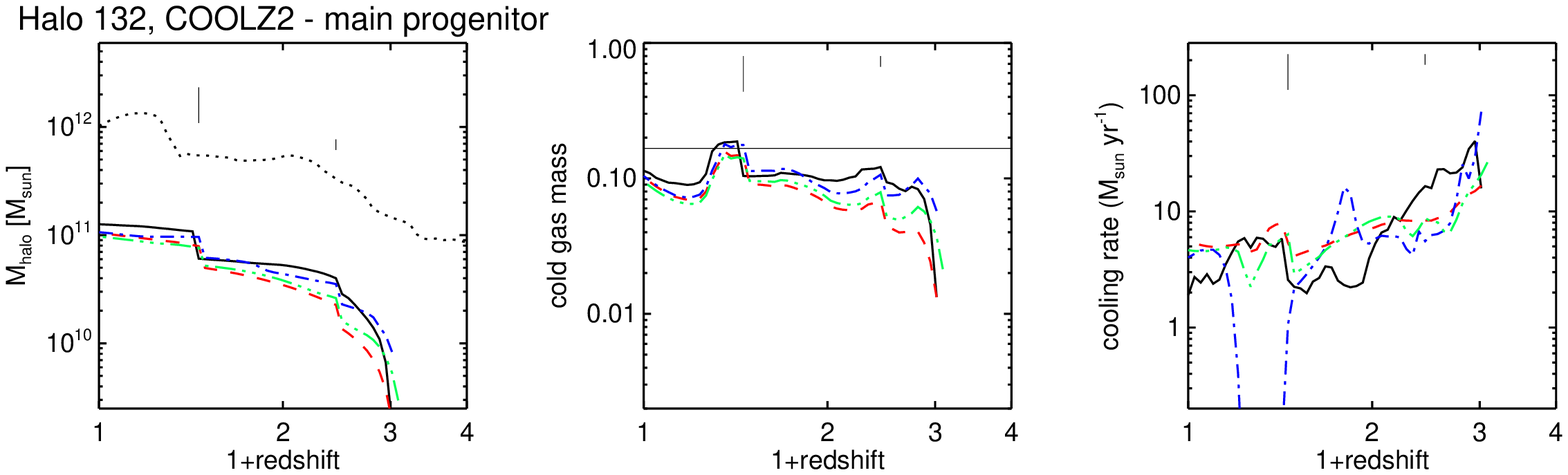}}
\centering{\includegraphics[width=.95\textwidth]{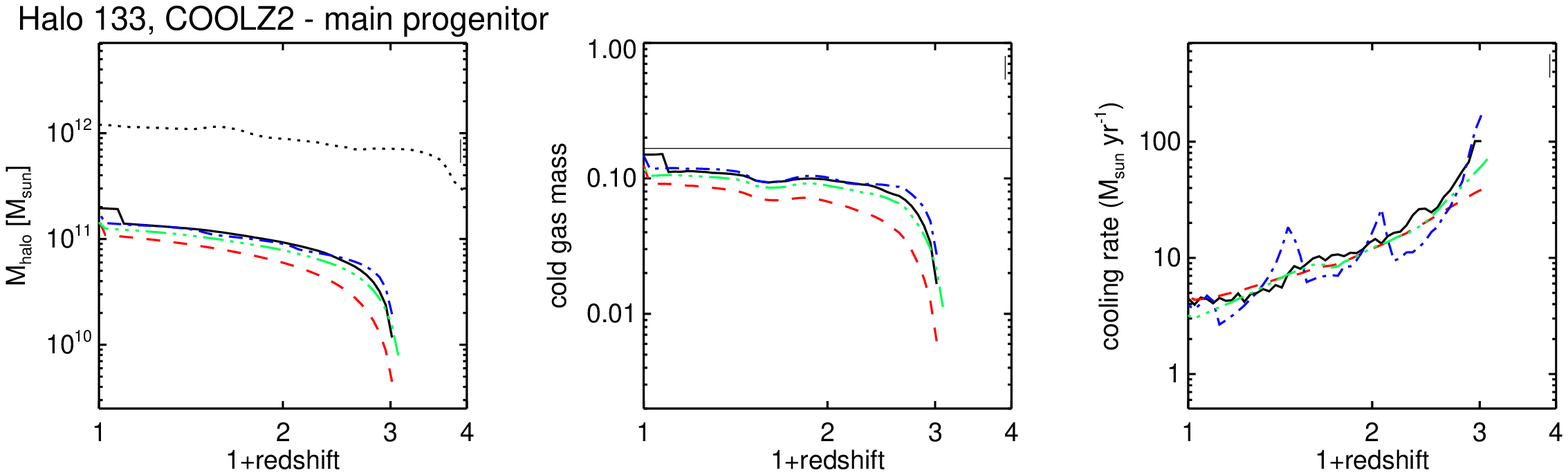}}
\caption{The same as Fig.~\ref{fig:trees} for the COOLZ2 simulation.}
\label{fig:treesz2}
\end{figure*}

\section{Results}
\label{section:results}

As in Paper I, we illustrate the behavior of models both by
considering the predicted cooled masses and cooling rates for all
halos more massive than $M_{\rm min\ halo}$ at some relevant
redshifts, and by showing detailed predictions for the four example
halos selected in Section~\ref{section:cooled}.  We recall that halos
0 and 132 experience major mergers below $z=1$, while halos 1 and 133
have more quiet merging histories.  To present a more organic
discussion, we show all the main results before discussing them.

Figs.~\ref{fig:scatters_cool} (COOL simulation) and
\ref{fig:scatters_z2} (COOLZ2 simulation) show, for each halo or its
main progenitor, a comparison between model predictions and simulation
results.  Starting from the upper panels, we give results at $z=0$ for
the three models separately; {\gal} is shown in red, {\mor} in blue,
{\del} in green.  For each set of $3\times2$ panels, the upper panels
report the cooled masses, the lower panels the cooling rates.  Pairs
of points corresponding to the same halo are connected by a thin black
line; this allows to highlight the cases in which SAM predictions lie
beyond the limits of the plot.  In all panels, the lower part reports
the ratio between SAM and simulation results, thick lines show median
values of the points in mass bins of 0.5 dex, while thin lines
correspond to the 16 and 84th percentiles computed as follows: calling
$R_i$ the ratio of model and simulation predictions for halo $i$, we
compute for each mass bin the median value $R_{\rm med}(M)$ and the
quantity $x_i=R_i/R_{\rm med}(M)$, then compute the 16 and 84th
percentiles of the set of $\{x_i\}$ for all halos and show these
percentiles scaled with the median value.
Results are given at $z=0$, $1$, $2$ and $3$ for COOL
(Fig.~\ref{fig:scatters_cool}), $z=0$ and $1.5$ for COOLZ2
(Fig.~\ref{fig:scatters_z2}).

The detailed evolution of the four example halos is shown in
Figs.~\ref{fig:trees} (COOL simulation) and \ref{fig:treesz2} (COOLZ2
simulation).  For each halo we show its mass accretion history
  for DM and cold gas, cold
gas fraction and cooling rate of the main progenitor.  Simulation
results are shown in black; color coding for the models is the same as
in the previous figures.

We confirm that cooling models embedded in SAMs are able to follow
relatively well the accumulation of cooled mass obtained in the
simulation.  As far as cooled masses are concerned, average
differences are smaller than $\sim$30 per cent, with some exceptions
that will be commented on later (Figs.~\ref{fig:scatters_cool} and
\ref{fig:scatters_z2}), and only a few halos show values that are
discrepant by more than a factor of two.  Cooled masses are also
recovered with a relatively small scatter, typically less than 20 per
cent.  Cooling rates are recovered in a much noisier way, with average
discrepancies and scatter being typically within a factor of two for
{\gal} and {\del}, within a factor of three for {\mor}.  The same
level of agreement is achieved in the detailed history of example
halos (Figs.~\ref{fig:trees} and \ref{fig:treesz2}).  Beyond this
overall agreement, a number of interesting discrepancies are evident.

\begin{figure}
\centering{\includegraphics[width=.45\textwidth]{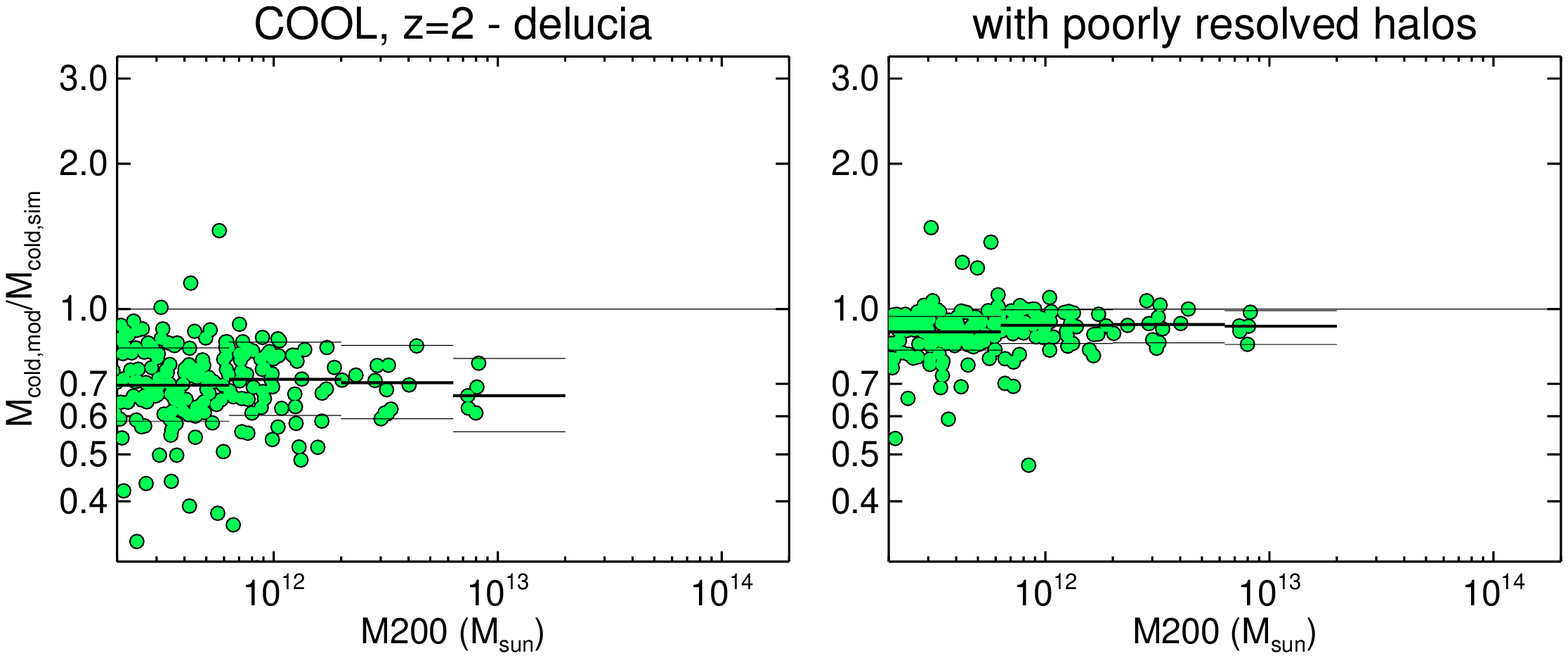}}
\caption{Ratio of cooled mass from the {\del} model and from the COOL
  simulation at $z=0$, as a function of halo mass.  Left panel:
  cooling on poorly resolved progenitors (less massive than $M_{\rm
    min\ prog}$) is subtracted out.  Right panel: all progenitors are
  considered.}
\label{fig:nomassive}
\end{figure}

(i) Cooled mass and cooling rates predicted by SAMs are typically
below the corresponding results from the COOL simulation at $z\ga2$.
This is apparent both in the histories of the four example halos
(Fig.~\ref{fig:trees}) and in the $z=2$ and $z=3$ panels of
Fig.~\ref{fig:scatters_cool}, where cooled masses are below
simulated ones by $\sim20-40$ per cent, while cooling rates are in
much better agreement.  This implies that models are slow in
accumulating cooled mass in the infall-dominated regime.  This result
is evident only when the mass cooled on unresolved halos is subtracted
out.  Fig.~\ref{fig:nomassive} shows how the cooled mass is recovered
by one of the models ({\del}) when the contribution from poorly
resolved halos is (left) or is not (right) subtracted out both in the
model and in the simulation.  Model results are in better agreement
with simulations in the second case, but this agreement is due to the
fact that the numerical underestimate of cooled gas happens to roughly
compensate the model underestimate.  Similar results are obtained for
the other two models.

(ii) For massive halos, the {\gal} model underestimates cooled masses
and cooling rate relative to simulations.  This is clearly visible for
halos 0 and 1, for which the {\gal} prediction of cooling rates drops
to zero at $z<1$.  Fig.~\ref{fig:scatters_cool} and
\ref{fig:scatters_z2} shows that at $z=0$, cooling rates for the most
massive halos ($M>2\times10^{14}\ {\rm M}_\odot$) are systematically
below simulation values and approach 0 in several cases.

(iii) {\mor} reproduces the simulated cooling rates with much more
scatter than the other two models: the fraction of halos discrepant by
more than a factor of 10 ranges from 13 to 20 per cent (10 to 15 per
cent for the other models). As apparent in Figs.~\ref{fig:trees} and
\ref{fig:treesz2}, this is connected to the behavior of the model at
major mergers: cooling is quenched for a period that corresponds to
$n_{\rm quench}$ dynamical times plus one cooling time, while no such
behavior is seen in simulations.  Cooling rates then peak to high
values and compensate the underestimate, thus providing on average
correct values for the cooled mass, though the scatter with respect to
simulated values is a higher than the other models.

(iv) At low redshift, both {\del} and {\mor} tend to overestimate
cooling rates in massive halos.  This is consistent with the finding
of \cite{Saro10} of higher cooling rates at low redshift in a galaxy
cluster.  In that paper this discrepancy was shown to be due to the
isothermal profile assumed for the gas in the {\del} model.  The same
trend is, however, present at very similar level in {\mor}, that
assumes a hydrostatic density profile with a shallow inner slope
closer to the actual gas profile found in the simulations. We will
come back to this issue later.

(v) Switching on cooling at $z=2$ has the effect of triggering quick
deposition of cooled mass and high cooling rates.  From
Fig.~\ref{fig:treesz2} we see that, while {\del} and especially {\mor}
follow this transient well and quickly converge to the cooled mass of
the simulation, {\gal} predicts lower cooling rates and is slower in
converging to the correct gas fraction; in massive halos, it always
remains below the simulation value.  Looking at
Fig.~\ref{fig:scatters_z2}, we see that {\mor} gives a good fit of
simulated cooled fractions already at $z=1.5$, though with substantial
scatter, where {\del} is biased low by $\sim30$ per cent and {\gal} by
a factor of two.

\subsection{The role of gas profile}
\label{section:profile}

\begin{figure*}
\centering{\includegraphics[width=.95\textwidth]{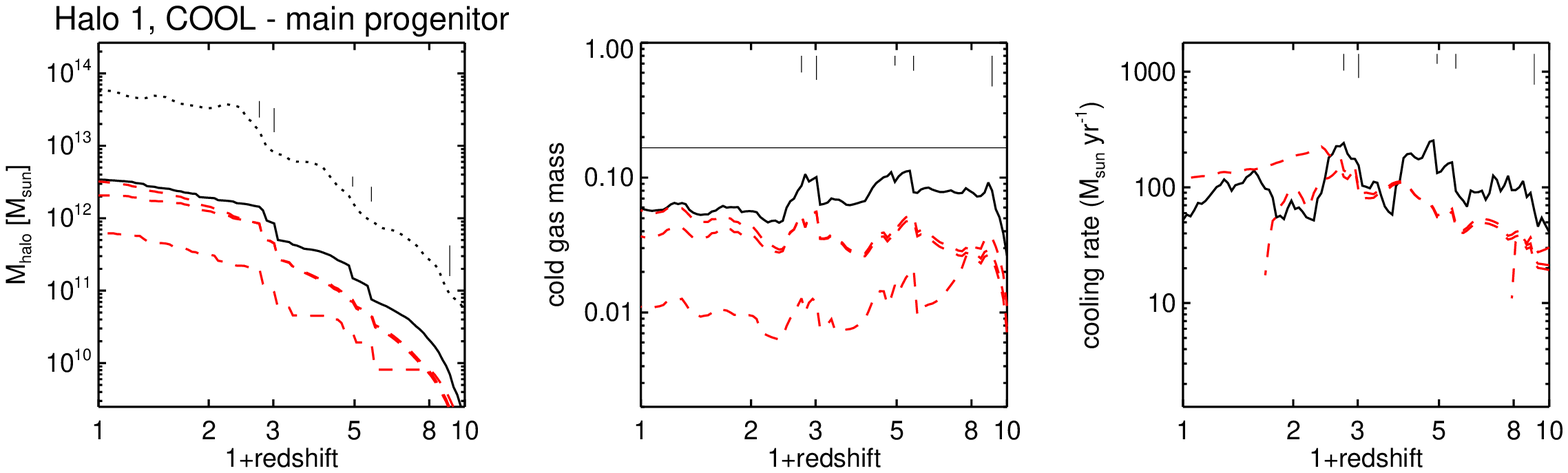}}
\centering{\includegraphics[width=.95\textwidth]{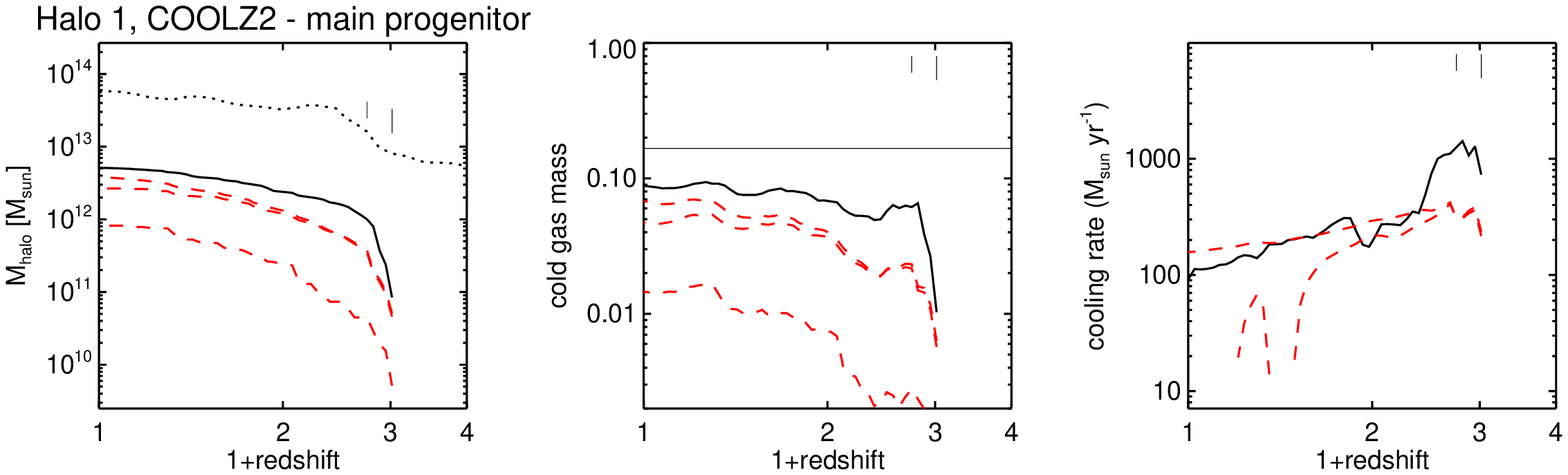}}
\caption{Results of the {\gal} model with different gas density
  profiles. The upper and lower rows give (as in Figs.~\ref{fig:trees}
  and \ref{fig:treesz2}), for the COOL and COOLZ2 simulation, the
  evolution of halo 1 for standard core (continuous line), isothermal
  (dotted line) and large core (dashed line) gas profiles. }
\label{fig:galacticus1}
\end{figure*}

\begin{figure*}
\centering{\includegraphics[width=.95\textwidth]{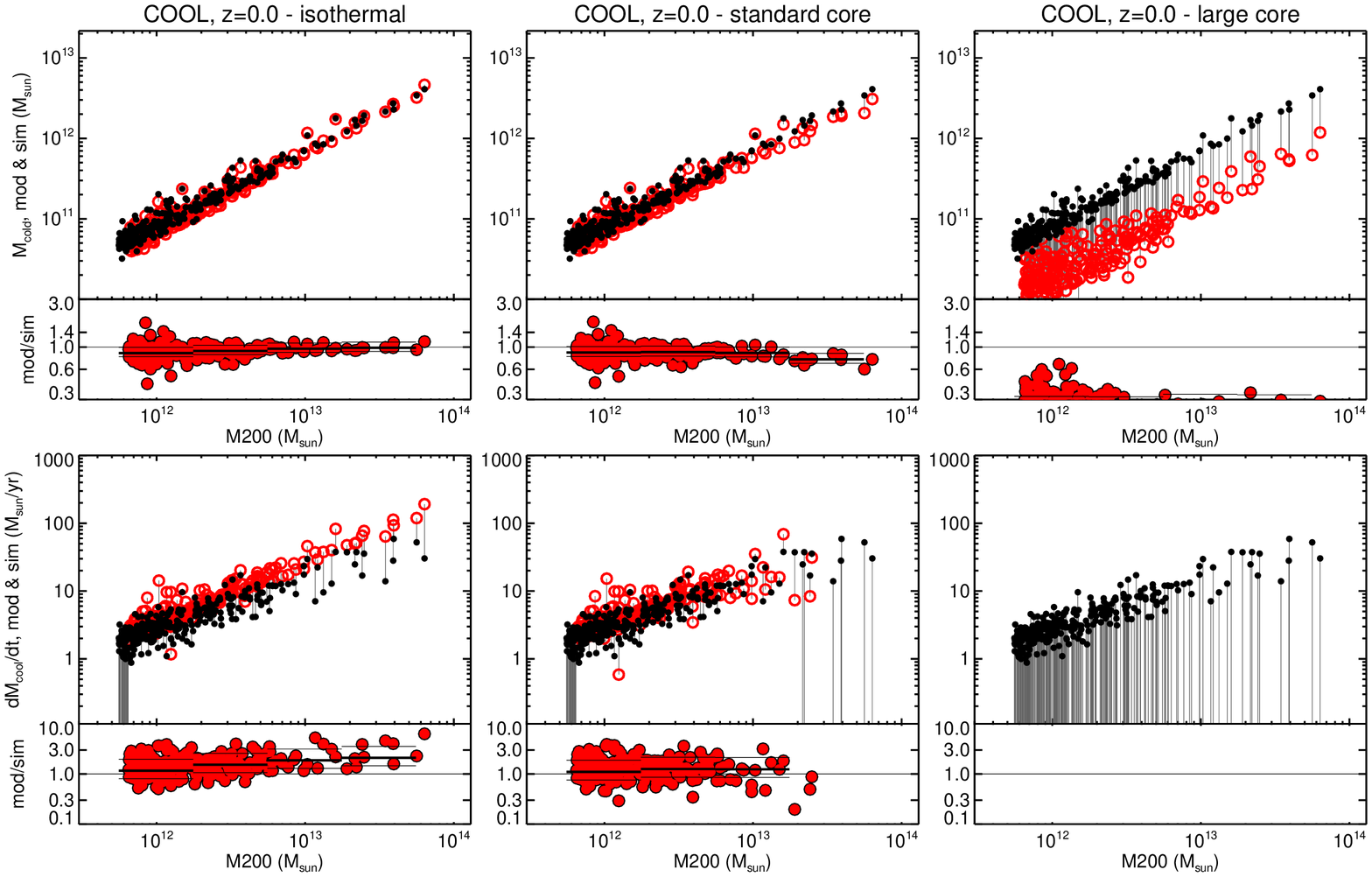}}
\centering{\includegraphics[width=.95\textwidth]{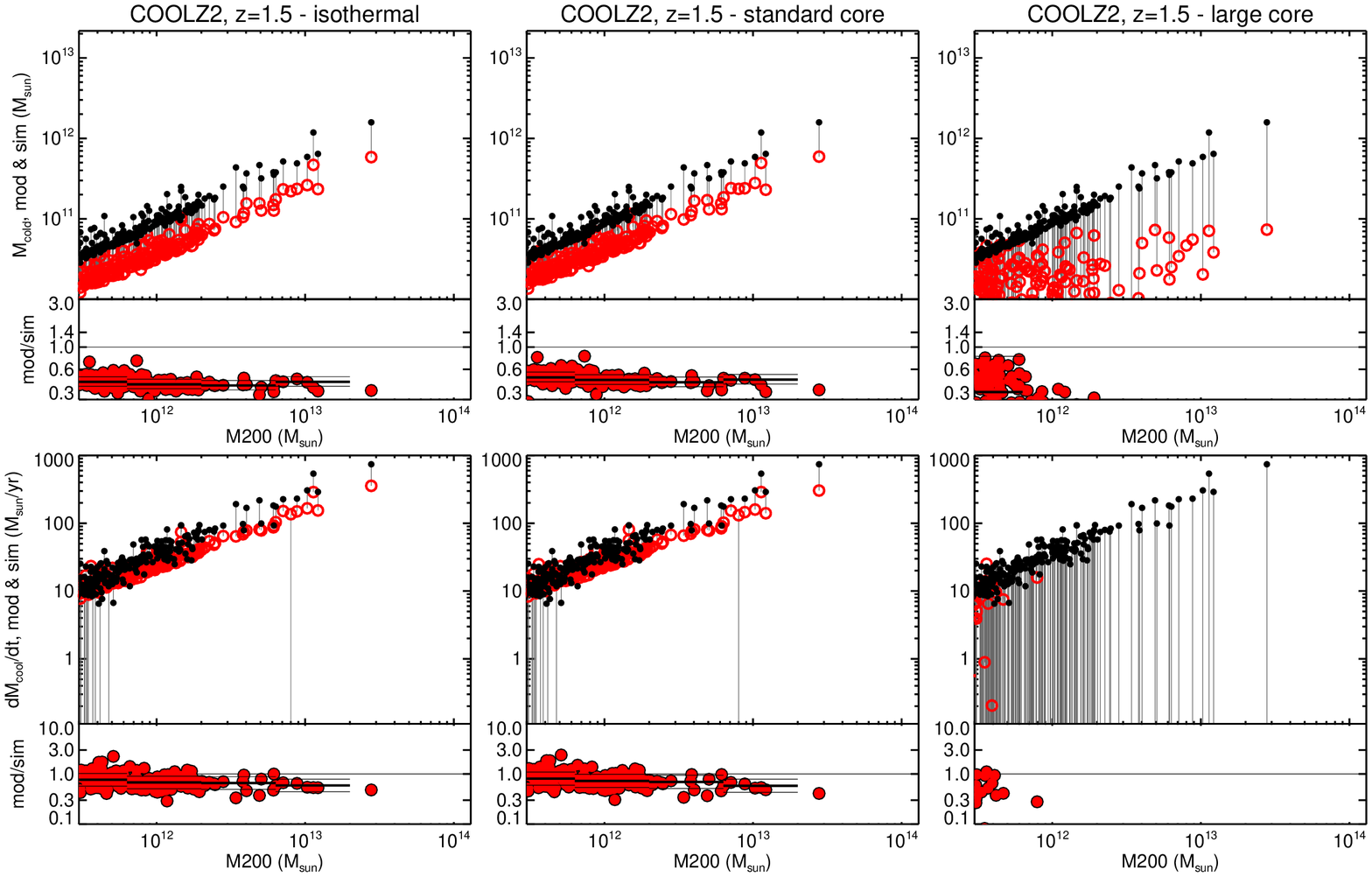}}
\caption{Results of the {\gal} model with different gas density
  profiles. The upper and lower panels give (as in
  Fig.~\ref{fig:scatters_cool}) cooled masses and cooling rates for all
  halos.  Upper panels give the results of the COOL simulation at
  $z=0$, lower panels those of the COOLZ2 simulation at $z=1.5$.  
  Left, mid and right panels refer to the {\gal} model with isothermal
  profile, standard core radius and large core radius.}
\label{fig:galacticus2}
\end{figure*}

Point (ii) is in line with the finding in Paper I that the cored gas
profile assumed in the durham model is mainly responsible for the
under-prediction of cooling rate with respect to the {\del} model,
that assumes an isothermal gas density profile.  In paper I, we
verified that, when a similar assumption is made for the gas profile
in the durham model, results from the two SAMs are much closer.  Point
(v) is in line with the results of \cite{Viola08}, who found that
their implementation of ``classical cooling'', very similar to the
cooling model of {\gal} for the case of a static halo, underestimates
cooling rates when cooling is suddenly switched on.  In that case the
gas profile was computed assuming a hot atmosphere in hydrostatic
equilibrium (as in {\mor}, Eqs.~\ref{eq:rho_mor} and \ref{eq:T_mor}),
and the resulting profile gave a good fit to the simulated one before
cooling was switched on.

To understand the influence of the assumption of gas profiles in the
cooling model, we run the {\gal} model with a singular isothermal
profile or a cored profile with a much larger core radius, of 0.42
times the virial radius.  Figs.~\ref{fig:galacticus1} and
\ref{fig:galacticus2} show some relevant results.  As an example for
this case, we chose halo 1, which has a more quiet merging history.
Fig.~\ref{fig:galacticus1} shows for the COOL and COOLZ2 simulations
the evolution of halo 1 for the three {\gal} models,
Fig.~\ref{fig:galacticus2} shows cooled masses and cooling rates for
all halos; for the COOL simulation, we show results at $z=0$, while
for COOLZ2, we show results at $z=1.5$ (recall that cooling was
switched on at $z=2$ in this run).  In agreement with results
discussed in Paper I, we find that using an isothermal gas density
profile brings SAM predictions in much better agreement with results
from the COOL simulation at low redshift.  In this case, significant
cooling rates are obtained in large halos at $z=0$, while results are
very similar for smaller halos.  In fact, this model behaves very
similarly to {\del}, which assumes the same gas density profile; this
similarity includes the tendency of cooling rates higher than
simulated ones in massive halos.

However, when the COOLZ2 simulation is considered, the assumption of
an isothermal gas density does not increase significantly the fraction
of cold gas at late times (Fig.~\ref{fig:galacticus1}).  As visible in
the lower panels of Fig.~\ref{fig:galacticus2}, cooled masses at
$z=1.5$ are as low (50 per cent lower than the simulation) as in the
standard core model, while the {\del} model is lower by 30 per cent
only (\ref{fig:scatters_z2}).  This is in line with \cite{Viola08},
where it was shown that the slow accumulation of cooled mass in the
``classical model'' is due to the assumption, implicit in
Eq.~\ref{eq:coolingrate}, that each gas shell cools on a cooling time
computed using the initial density and temperature, while the
evolution of gas elements is at increasing density and roughly
constant temperature, and this leads to catastrophic cooling on a
shorter time scale.  The different results of {\del} and {\gal} show
that the result is sensitive to the precise integration scheme of
Eq.~\ref{eq:coolingrate}, however.

The results discussed above show that simulations disfavour the use of
a very large core for the gas profile in the {\gal} model, at least in
this setting where gas density profile is not affected by any source
of feedback.  Clearly, feedback is going to influence gas profile and
may increase the core size but, whenever the halo is well within the
cooling-dominated regime, feedback from star formation will be
triggered by cooling itself, so it would be difficult to sustain a
very large core radius if this has such a large negative impact on the
cooling rate.

Another parameter related to the halo density profile is the
polytropic index $\gamma_p$ of the {\mor} cooling model
(Eqs.~\ref{eq:rho_mor} and \ref{eq:T_mor}).  Because this quantity is
constrained both by simulations and, for galaxy clusters, by
observations, there is no much freedom to vary it.  We ran the model
assuming $\gamma_p=1.05$ and $1.25$, and found that the two new
realizations were giving slightly larger discrepancies with respect to
the simulations, consistently with $1.15$ being the optimal value.

\subsection{Behavior during major mergers}
\label{section:majormergers}

One reason why {\mor} produces a large scatter in the predicted
cooling rates is that it assumes that cooling is switched off in major
mergers.  The rationale behind this idea is the deep reshuffling of
phase space that takes place at these events.  Indeed, looking at
Figs.~\ref{fig:trees} and \ref{fig:treesz2}, no obvious decrease in
cooling rates is visible at or after major mergers in simulations.  We
looked for correlations between the ratio of progenitor masses in two
consecutive outputs and the ratio of cooling rates before the merger
and after $n$ outputs, with $n$ varying from 1 to a few.  The first
quantity is large when a major merger takes place, so if major mergers
quench cooling then we would expect the second quantity to be
negatively correlated with the first one.  We did not find any such
correlation.

The lack of any quenching of cooling at major mergers is plausibly
connected to the persistence of strong condensations at the center of
DM halos.  In the case of runaway cooling studied here, the resulting
condensations deepen the halo potential well and thus cause an
enhancement of hot gas density and then of radiative cooling around
them.  In other words, they act as seeds for cooling of hot gas.
These condensations are simulated as clumps of collisionless stars,
and are so compact that large-scale tides are ineffective at
disrupting them.  The effect of merging-induced shock waves on the hot
gas that collects around such condensations will be to some extent
affected by the hydro solver, but this should be a further-order
effect.  It would be very interesting to test this idea with a
Eulerian simulation that uses the same setting.

Conversely, when the whole physics of galaxy formation is taken into
account, ``condensations'' will not represent unphysical condensations
but true galaxies.  In this case, further heating will be provided by
energetic feedback from massive stars and AGN.  This heating source
will be able to limit the accumulation of mass in galaxies and to
flatten the density profile of hot gas around them, keeping it to a
higher adiabat and making it easier to heat.  In these conditions,
major mergers may still cause a quenching of cooling.  Of course AGN
feedback would likely be more effective in this regard.

\begin{figure}
\centering{\includegraphics[width=.45\textwidth]{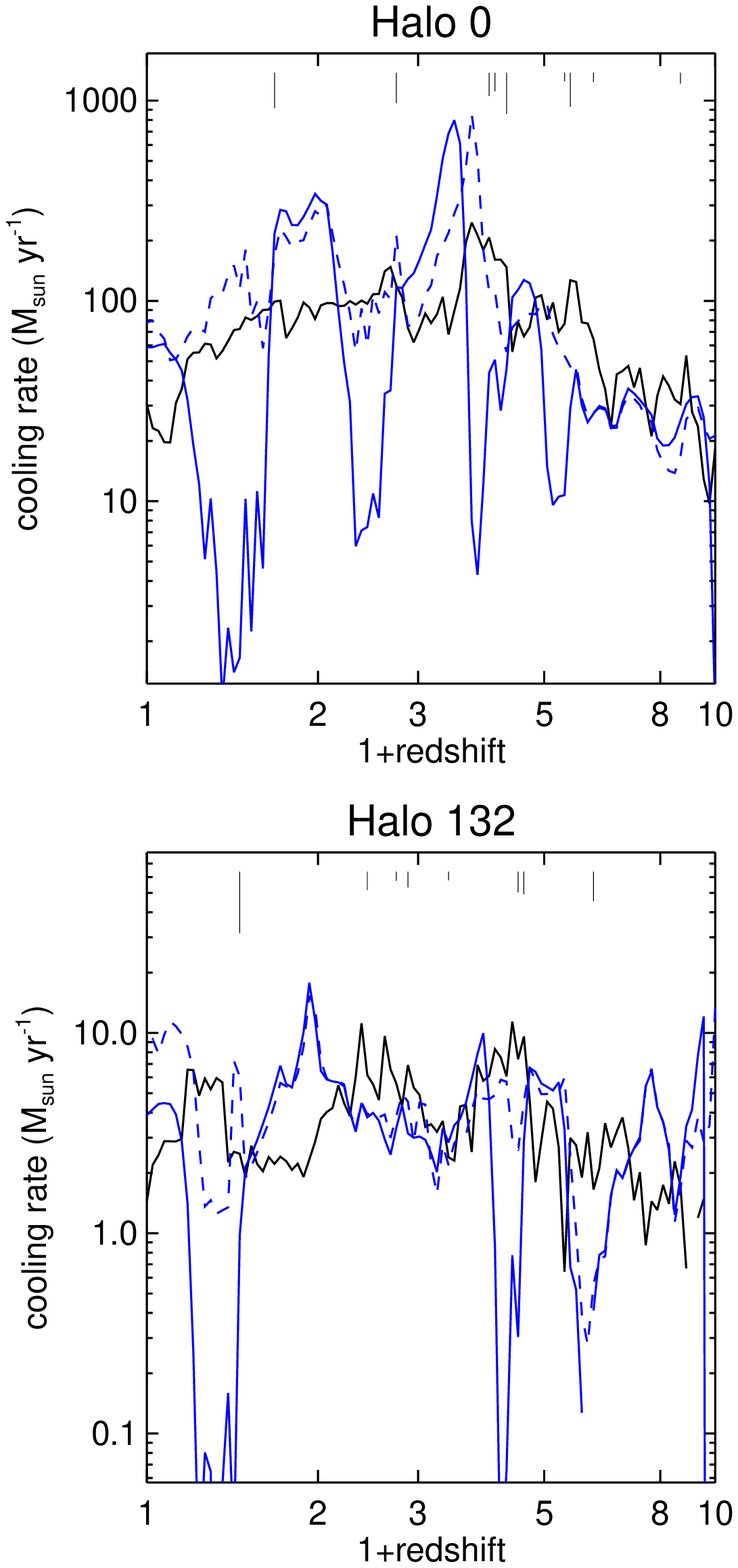}}
\caption{Evolution of cooling rates in the COOL simulation for the two
  example halos with late mergers (0 and 132). Results of the {\mor}
  model are shown for two different assumptions on the behavior at
  major mergers. Continuous line: standard cooling model; dashed
  lines: cooling is not quenched at major mergers.}
\label{fig:morgana1}
\end{figure}

\begin{figure}
\centering{\includegraphics[width=.45\textwidth]{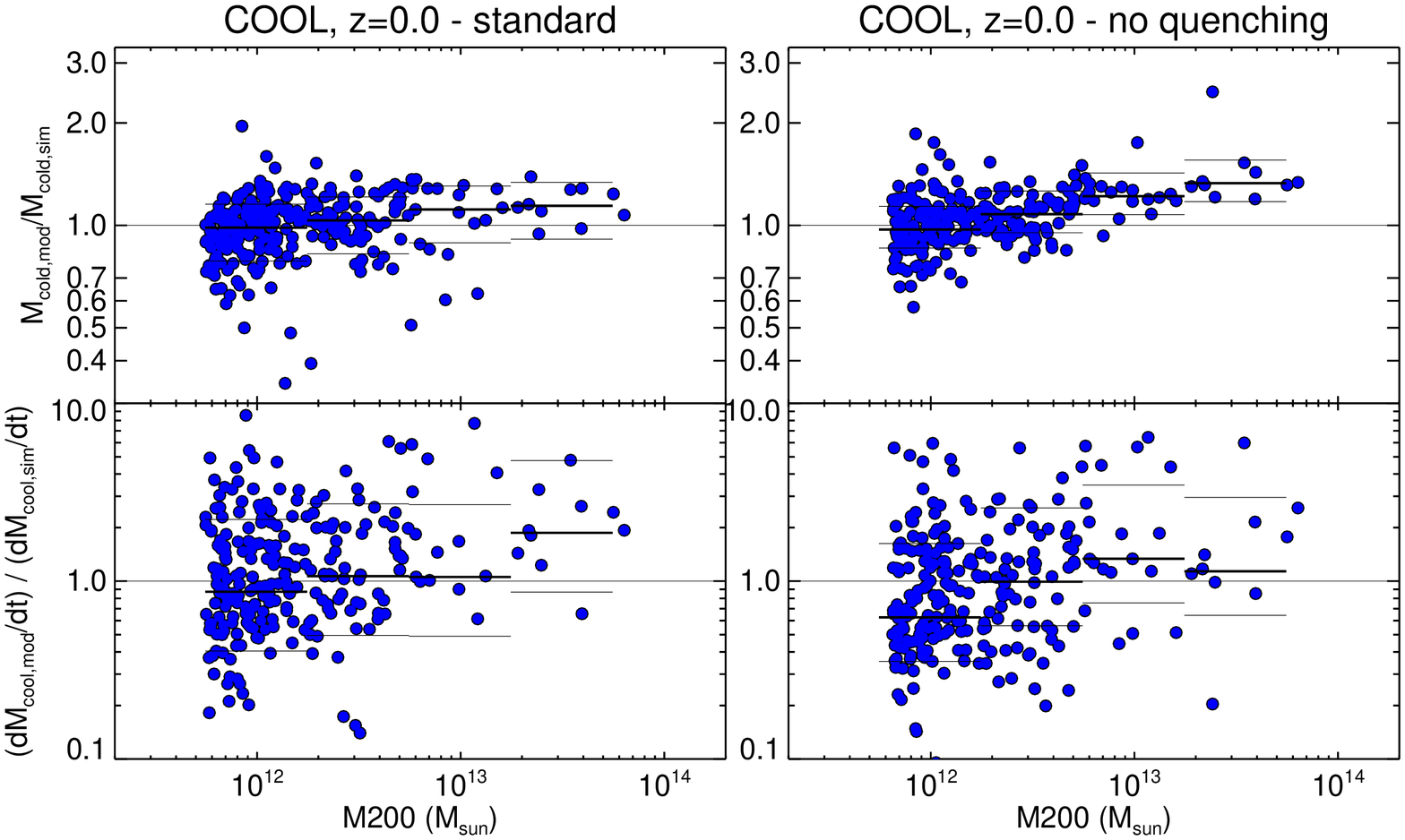}}
\caption{Cooled masses and cooling rates for all halos in the COOL
  simulation at $z=0$.  Results of the {\mor} model are shown for two
  different assumptions on the behavior at major mergers. Left
  panels: standard cooling model; right panels: cooling is not
  quenched at major mergers.}
\label{fig:morgana2}
\end{figure}

It is interesting to see what happens to {\mor} when quenching of
cooling at major mergers is not applied.  Fig.~\ref{fig:morgana1}
shows, for the COOL simulation, the cooling rates of the two example
halos that suffer major mergers at late times (0 and 132).  The
continuous line shows the standard cooling model, while the dashed
line shows a model with no quenching.  Here the cooling radius is
still reset to zero at major mergers; we have verified that dropping
this condition does not lead to significant changes.  The model
without quenching removes some troughs in the cooling rates.  In
Fig.~\ref{fig:morgana2} we show the cooled gas and the cooling rates
predicted with the two models for all halos.  Left and right panels
give the results of the standard and no quenching model respectively.
As expected, the number of halos where cooling rates differ from the
simulation values by more than a factor of 10 is less in the no
quenching model, but several cases are still present and the scatter
does not decrease significantly.  Because of the systematically higher
cooling rates, the cooled masses are now biased high by $\sim10$ per
cent at high masses.

One more possibility to investigate is to remove the term in
Eq.~\ref{eq:coolingradius} that decreases the cooling radius at the
sound speed.  This term is peculiar to the {\mor} cooling model, where
the cooling radius is treated as a dynamical variable.  This term
helps the gas to drift towards the halo center, so it causes higher
cooling rates at late times after a reset of the cooling radius.  We
have verified that if we drop this term in the model with no quenching
we obtain slightly lower cooling rates, that limit the overestimate of
cooled mass with respect to the simulation, but the resulting cooled
masses are biased low by $\sim$30 per cent at low masses.

\section{Comparison with previous results}
\label{section:comparison}

With respect to previous papers that performed similar tests based on
SPH \citep{Benson01,Yoshida02,Helly03,Cattaneo07,Lu11}, our
simulations improve significantly in terms of mass and force
resolution. \cite{Hirschmann12} worked at a resolution and in a mass
range analogous to the one used in this paper, while \cite{Saro10},
presented results for a galaxy cluster at a resolution that is
analogous to ours for that mass scale.  Together with our good time
sampling, good mass resolution has allowed us to limit the computation
of cooled masses to those halos that are resolved with more than
$\sim10000$ particles.  Our conservative limit is much higher than
what has commonly been used, and guarantees a proper sampling of the
central region where cooling takes place.  This high value of $M_{\rm
  min\ prog}$ also removes the contribution to cooling from those
halos that are affected by the UV background.  As an example of the
importance of this effect, we showed (Figure~\ref{fig:nomassive}) that
the underprediction of the amount of cold baryons in the
infall-dominated regime (also reported in other papers, see below),
can be compensated for by the effect of poor resolution when much
cooling takes place at high redshift.

A detailed comparison of our results to the most recent papers
\citep{Saro10,Lu11,Hirschmann12} is complicated by the different
treatment of merging times and resolution effects.  All papers agree
in identifying, beyond a general broad agreement, a trend of
underestimation of cooled masses at high redshift or small halo mass
and overestimation at low redshift or high halo mass, trends that we
find also in our analysis.  The high-redshift/low mass trend was
explained by \cite{Lu11} and \cite{Hirschmann12} as an effect of the
more extended dominance of cooling-dominated regime with respect to
SPH simulations, where cold flows are found to easily penetrate the
hot halos and quickly deposit their mass into the central galaxy.  The
low-redshift overestimation was interpreted by \cite{Saro10} as an
effect of their assumption of a singular isothermal profile for the
hot gas, that is at variance with the flatter profile found in
simulations.  To this last conclusion we can add that the assumption
of a cored profile in {\gal} leads to an underestimation of cooled
masses, while the assumption of hydrostatic equilibrium profile in
{\mor} leads to a similar overestimation. This shows how the details
of the implementation are of great importance.

The only analogue of our COOLZ2 simulation is the simulation of static
halos performed by \cite{Viola08}.  Our results are fully consistent
with those presented in that paper and they extend the same
conclusions in the much more realistic environment of full dark matter
merger trees: following the quick onset of cooling is a challenge for
cooling models.  The poor performance of a model, like {\gal}, based
on \cite{Cole00} was explained in that paper by showing that
Eq.~\ref{eq:coolingrate} is strictly valid as long as a mass element
cools to low temperatures in a time equal to the cooling time computed
using its initial density and temperature, but gas elements evolve at
roughly fixed temperature and increasing density, and thus cool more
quickly.  While all models use some version of
Eq.~\ref{eq:coolingrate} for computing the cooling rate, the specific
implementations makes the results very different for this test.

The difference between {\sc gadget} and the moving-mesh code {\sc
  arepo}, reported by \cite{Keres12} and \cite{Nelson13}, raises the
worry that our results may be affected by the use of a specific hydro
solver.  Again, the results presented in \cite{Keres12} refer to the
central galaxy and make no attempt to subtract out the contribution of
poorly resolved halos, so a straightforward comparison is not
possible.  However, we computed at $z=0$ the mass of central cooled
gas concentrations (using the ``star'' particles within $1/10$ of the
virial radius) and compared it with the galaxy mass/halo mass relation
of \cite{Keres12} (their figure 4), finding within the statistics a
good agreement with the SPH result of that paper.  We also checked
density and temperature profiles of halos, and found them to be
roughly consistent with what presented in Figures 8 and 9 of that
paper, and inconsistent with the drop in temperature found in {\sc
  arepo} halos.  As expected, despite the different ``star formation''
algorithms used, and in absence of effective thermal feedback, we
obtain very similar results when we use the same code.

Intriguingly, both {\sc arepo} and SAMs predict, with respect to SPH,
higher cooled masses and cooling rates at low redshift/large halo
mass.  We find that SAMs overpredict cooling rates by a factor of
$\sim2$ and cooled masses by a much smaller factor, while
\cite{Keres12} report much larger differences between the two codes.
However, for a proper comparison one should check what is the
difference when merging is neglected and the contribution from poorly
resolved halos is subtracted out.  Regardless, a conclusive assessment
of the accuracy of different implementations of hydrodynamics is
necessary before firm conclusions on the behavior of cooling in
massive halos can be reached.

\section{Summary and discussion}
\label{section:discussion}

We have tested cooling models embedded in three widely used SAMs
\citep{galacticus,Monaco07,DeLucia07} by comparing their predictions
to N-body hydrodynamical (SPH) simulations of radiative cooling in
cosmological DM halos.  With respect to previous papers that performed
similar tests
\citep{Benson01,Yoshida02,Helly03,Cattaneo07,Viola08,Saro10,Lu11,Hirschmann12},
our simulations are improved in several ways: (i) we used a mass and
force resolution sufficient to fully resolve the cooling region in all
halos larger than $5\times10^{10}\ {\rm M}_\odot$; (ii) we computed
cooling rates and cooled masses subtracting out the contribution of
cooling in poorly resolved halos; (iii) by also running a simulation
where cooling is switched on at $z\sim2$, we were able to test cooling
models exactly in the redshift range where they are expected to be
valid, with no influence from high-redshift over-cooling; (iv) using a
suitable formulation of SPH and a ``star formation'' algorithm
\citep[as in][]{Viola08,Saro10} to treat cooled particles as
collisionless, we were able to limit numerical cooling and speed up
the simulation considerably.  As a word of caution, we recall that the
simulations that we compare to SAMS have been carried out using an SPH
hydrodynamical solver.  Significant differences have been reported for
the amount of star formation predicted by a simulation based on the
{\sc arepo} code that uses an Eulerian scheme
\citep{Keres12,Nelson13}, so some of our results, especially those
relative to massive halos, may be affected by the specific hydro
solver we use.

We confirm that, overall, cooling models are able to approximately
predict the correct amount of cooled mass. When cooling is active
since the start of the simulation (the COOL simulation), median values
are recovered at $z=0$ to within 20 per cent and with a similar
scatter.  At higher redshift, median values agree to within 40 per
cent.  For individual halos, the worst discrepancies are rarely larger
than a factor of two.  However, when the cooled mass in the COOL
simulation is limited to properly resolved halos, cooling models in
SAMs underestimate the amount of cooled mass in the infall-dominated
regime by $\sim20-40$ per cent at $z=2$.  This difference is much less
visible when the contribution of poorly resolved halos is included, as
the numerical underestimate happens to compensate the theoretical one.

In some models ({\del} and {\mor}, and {\gal} when an isothermal gas
profile is used), cooled masses for more massive halos tend to be
higher than those found in the simulations at $z=0$. 
While
 this difference is found in the same range of halo masses
  where Eulerian and Lagrangian codes give different results \citep{Keres12}, this
confirms the findings of \cite{Saro10} (see also \citet{Hirschmann12}), who compared predictions from
the {\del} model with cooling rates from a simulation with very
similar setting as our COOL simulation.  This is a very relevant
point: late cooling in massive halos deposits mass in the central
galaxy, which is typically a bright elliptical galaxy.  The stellar
populations of these galaxies are observed to be very old
\citep[e.g.][]{Thomas05}, so quenching this cooling is necessary for
any successful galaxy formation model.  AGN feedback in the so-called
radio mode \citep{Croton06,Bower06} it typically advocated to produce
this quenching; however, the fraction of galaxies with detectable
radio emission associated with the AGN activity necessary to perform
this quenching is higher than what it is observed and shows the wrong
dependence as a function of halo mass \citep{Fontanot11}.  As noticed
also in Paper I, a cooling model that produces too strong a cooling
flow at late times would require stronger AGN feedback to maintain
quenching.  For the {\del} model, \cite{Saro10} showed that the higher
cooling rates with respect to results from simulations are due to the
assumption of isothermal gas density profile.  The same trend is shown
by {\mor}, which assumes a hydrostatic density profile with a shallow
inner slope.  Clearly, the different type of integration used in the
{\mor} cooling model causes the same trend without assuming a singular
profile.  
However, before reaching a firm conclusion on this point it is
necessary to fully understand the role of the hydrodynamic scheme.

Consistently with the results of Paper I, the {\gal} model with a cored
gas profile under-predicts cooling flows in massive halos at late
times with respect to simulations.  This prediction depends
significantly on the assumed gas density profile: using an isothermal
gas density profile generates predictions that are very similar to the
{\del} model and in much better agreement with the simulation, while
large core radii are strongly disfavoured.

Cooling rates found in the simulations are recovered in the models
with larger scatter, about a factor of two.  Scatter is larger for the
{\mor} model, and part of it is due to the assumption in {\mor} that
cooling is quenched during major mergers.  We find no such trend in
the simulations, and this is likely due to the persistence of cooled
condensations during mergers.  When the assumption of quenching at
major mergers is dropped, {\mor} predicts cooling rates with less
scatter, but it still larger than the other models.

Comparing models and simulations when cooling is switched on at $z
\sim 2$ probably represents the cleanest test of the cooling model.
In this case, in both the simulation and models, the baryons
associated with halos are not pre-processed by runaway cooling, they
are all located in the hot atmospheres when cooling is switched
on. Also, the contribution of cooling from poorly resolved halos is
very small for several halo dynamical times, and the deposition of
cold gas is much less affected by all the numerical issues connected
with the infall-dominated regimes.  In our COOLZ2 simulation, the
cooled mass is found to accumulate rapidly. At $z=1.5$, {\mor} is able
to predict the right amount of cooled mass, though with substantial
scatter; {\del} is biased low by $\sim20-30$ per cent; and {\gal} is
low by a factor of 2, even when an isothermal gas profile is assumed.
This confirms the findings of \cite{Viola08}, obtained with static
halos and using gas profiles that closely matched the simulated ones
just before cooling is switched on.

Although switching on cooling at a given redshift provides a clean
test of cooling models, one can wonder whether such conditions occur
in realistic cases, and therefore, whether the difference among
models, which is much less evident in the COOL simulation, should be a
cause for concern.  When cooling is active since the beginning, most
baryons have already cooled by $z=2$ (especially in relatively small
halos).  The effect of this overcooling is visible, for instance, in
Fig.~\ref{fig:trees}, where cooling rates in the example halos 0 and 1
hardly exceed the value of $\sim200$ M$_\odot$ yr$^{-1}$.  These
deposition rates can translate at best into star formation rates of
the same order, which would be typically lower than the several
hundreds M$_\odot$ yr$^{-1}$ measured for massive star forming
galaxies at the same redshift.  In the COOLZ2 simulation, cooling
rates are higher by a factor of $\sim5$ when cooling is suddenly
switched on.  Feedback from massive stars and accreting black holes is
responsible for limiting overcooling at high redshift, and this makes
more gas available at lower redshift.  Moreover, as commented in
Section~\ref{section:majormergers}, halo mergers or, more likely,
feedback from star formation and AGN will likely be responsible for
episodic quenching of catastrophic cooling, so the setting of the
COOLZ2 simulation may be a good approximation for cooling flows in
massive halos at the peak of cosmic star formation, after a quenching
event.

This is the first time that several cooling models run on the same
merger trees are compared with a cosmological hydrodynamical
simulation.  Despite the simplified setting used in these simulations,
we believe that they provide an important benchmark test for cooling
models embedded in SAMs.  Indeed, while energetic feedback from stars
and AGN, which almost certainly plays a crucial role in shaping the
properties of galaxies, is the most important contributor to the
variance among model predictions \citep{Fontanot13}, an accurate
calibration of the condensation of gas in the central galaxy is
desirable to remove unwanted sources of inaccuracies.  Merger trees
and results from the simulations presented in this paper are available
to interested modelers upon request.

\section*{Acknowledgments}

We thank Giuseppe Murante for his help in running the simulations and
for discussions. We thank Volker Springel, Richard Bower and Michaela Hirschmann for their
feedback on this paper.
Simulations were run on SP6 at CINECA thanks to the
ISCRA-B project ``Radiative cooling in galaxy formation: a numerical
benchmark test for galaxy formation models''.  P.M. and
S.B. acknowledge financial contributions from the European Commissions
FP7 Marie Curie Initial Training Network CosmoComp
(PITN-GA-2009-238356), from PRIN MIUR 2010-2011 J91J12000450001 ``The
dark Universe and the cosmic evolution of baryons: from current
surveys to Euclid'', from PRIN-INAF 2009 ``Towards an Italian Network
for Computational Cosmology'', from ASI/INAF agreement I/023/12/0,
from PRIN-MIUR09 ``Tracing the growth of structures in the Universe''
and from FRA2009 and FRA2012 grants of the Trieste University.  GDL acknowledges
financial support from the European Research Council under the
European Community's Seventh Framework Programme (FP7/2007-2013)/ERC
grant agreement n. 202781.  FF acknowledges financial support from the
Klaus Tschira Foundation and the Deutsche Forschungsgemeinschaft
through Transregio 33, ``The Dark Universe''.

\bsp

\label{lastpage}

\bibliography{master}

\end{document}